\documentclass[thmsa]{article}
\usepackage{amsmath}
\usepackage{amsfonts}
\usepackage{amssymb}
\def\proof{\noindent\textsc{Proof. }}
\def\abstract{   \if@twocolumn
     \section*{Abstract} 
   \else 
     \small 
     \quotation 
     {\sc Abstract. } 
   \fi
}
\def\endabstract{\if@twocolumn\else\endquotation\fi}
\evensidemargin=\oddsidemargin
\newdimen\dummy
\dummy=\oddsidemargin
\def\tighttoc{\def\l@section{\@dottedtocline{1}{0em}{1.4em}}}
\tighttoc
\newtheorem{theorem}{Theorem}
\newtheorem{corollary}[theorem]{Corollary}

\newtheorem{lemma}[theorem]{Lemma}

\newtheorem{definition}[theorem]{Definition}

\begin{document}

\title{Twist Positivity for Lagrangian Symmetries\footnote{Work supported in part by
the Department of Energy under Grant DE-FG02-94ER-25228. This research was
carried out in part for the Clay Mathematics Institute.}}
\author{Olivier Grandjean, Arthur Jaffe, and Jon Tyson\\Harvard University \\Cambridge, MA 02138, USA}
\date{July 13, 2000}
\maketitle
\begin{abstract}
We prove twist positivity and positivity of the pair correlation function for
combined spatial and internal symmetries of free bosonic Lagrangians. We work
in a general setting, extending the results obtained in \textit{Twist
Positivity} [\ref{TwistPositivity}].
\end{abstract}

\thispagestyle{empty}
\newpage
\section{Introduction}
In this paper we generalize the results of {\it Twist Positivity} \cite{TwistPositivity} 
to a wide class of symmetries.  We investigate the (unitary) implementation $U_S$ of a 
symmetry $S$ of a classical, free field Lagrangian.  We establish twist 
positivity of the the partition function $Z_{\beta, U}$ twisted by the symmetry $U_S$, 
namely $Z_{\beta, U}>0$, and     
show positivity of the pair correlation operator that is  
twisted by $U_S$, namely $C_\beta>0$.  After considering a simple example of 
a free field in \S 1.1, we give the general definitions of these concepts in \S 1.2--1.3. 
The methods here are applicable both in quantum field theory and also in related
problems in statistical physics. 

\subsection{Free Bosonic Fields on Compact Manifolds.{\label{firstsect11}}}
Let $M$ be a
compact Riemannian manifold, $E\rightarrow M$ a Hermitian vector
bundle on $M$ endowed with a compatible connection, and
let $\Delta_E$ denote the corresponding (positive self-adjoint) Laplacian on the Hilbert space
$L^2(E)$ of square integrable sections of $E$.
Let $\left\langle \cdot,\cdot\right\rangle $ denote the
inner-product\footnote{Inner products are antilinear in the first argument.}
on $L^2(E)$, and let
$\mathcal{D}\left(  \triangle_{E}\right) $ be the domain of $\triangle
_{E}$. 

The corresponding free field theory has
the Lagrangian
${\cal L}: \mathcal{D}\left(  \triangle_{E}^{1/2}\right) \times L^2(E)\rightarrow {\mathbb R}$,
defined by,
\begin{equation}
\mathcal{L}\left(  \varphi_{\text{cl}},\frac{\partial\varphi_{\text{cl}}}{\partial t}\right)  =\left\langle \frac{\partial\varphi_{\text{cl}}}{\partial t},\frac{\partial\varphi_{\text{cl}}}{\partial t}\right\rangle
-\left\langle \triangle_{E}^{1/2}\varphi_{\text{cl}},\triangle_{E}^{1/2}\varphi_{\text{cl}}\right\rangle
-m^{2}\left\langle \varphi_{\text{cl}},\varphi_{\text{cl}}\right\rangle,
\end{equation}
where $m\geq 0$ is the \textit{mass} of the field.
The characteristic feature of free bosonic quantum field theory is that the time 
evolution is prescribed by a \emph{linear} partial differential equation of 
second order.  
The dynamics corresponding to this Lagrangian via the Euler
variational principle reads,
\begin{equation}
\left(  \frac{\partial^{2}}{\partial t^{2}}+\triangle_{E}+m^{2}\right)
\varphi_{\text{cl}}\left(  t,x\right)  =0 \ .
\label{KleinGordon}
\end{equation}
Since the manifold $M$ is compact, the self-adjoint operator
$\Delta_E$ has discrete spectrum and 
there is an orthonormal basis\footnote{See \S
$\ref{nobasissec}$ for a basis-independent quantization.} of
$L^2(E)$, $\{\varphi_{{\rm cl}, k}\}_{k\in K}$, consisting of 
eigensections of $\Delta_E$,   
\begin{equation}
\left(  \triangle_{E}+m^{2}\right)  \varphi_{\text{cl},k}=\omega_{k}^{2}\varphi_{\text{cl},k}\ .
\end{equation}

Restricting consideration to the case that all the $\omega_k^2$ are strictly positive,
each solution to eq.(\ref{KleinGordon}) may be expanded as
\begin{equation}
\varphi_{\rm cl}(t,x) =
\sum_{k\in K}\frac{1}{\sqrt{2\omega_k}}
(\bar{\alpha}_+(k) e^{i\omega_k t} +
\alpha_-(-k)e^{-i\omega_k t})\varphi_{{\rm cl}, k}(x)\ ,
\end{equation}
where $\alpha_\pm(\pm k)$ are complex coefficients.
Canonical quantization  of that system replaces
the complex coefficients $\alpha_\pm(\pm k)$ by operators, also
denoted by $\alpha_\pm(\pm k)$, satisfying the canonical commutation
relations,
\begin{eqnarray}
\left[\alpha_\pm(\pm k), \alpha^*_\pm(\pm k')\right] &=&
\delta_{k,k'}\nonumber\\ 
&&\label{CCR}\\
\left[\alpha_\pm(\pm k), \alpha_\pm(\pm k')\right] &=&
\left[\alpha_+(k), \alpha_-(-k')\right] 
= \left[\alpha_+(k), \alpha^*_-(-k')\right]=0\ .\nonumber
\end{eqnarray}
The $\alpha_{\pm}^{\ast}(\pm k)$ act on a Fock space $\frak{F}$, which is the Hilbert
space spanned by all vectors of the form $\alpha_{\pm}^{\ast}\left(  \pm
k_{1}\right)  ...\alpha_{\pm}^{\ast}\left( \pm k_{n}\right)  \left|
0\right\rangle $, where the unit vector $\left|  0\right\rangle $ (called the \textit{vacuum}) is
in the nullspace of all the $\alpha_{\pm}\left(  \pm k\right)  $
and in the domain of any product of $\alpha_{\pm}^{\ast}\left(  \pm k\right)  $'s.
The charged ``one-particle'' subspaces of $\frak{F}$, which are the spans of
$\left\{  \alpha_{-}^{\ast}\left(  -k\right)  \left|  0\right\rangle \right\}
_{k\in K}$  and  $\left\{  \alpha_{+}^{\ast}\left(  k\right)  \left|  0\right\rangle
\right\}  _{k\in K}$, play a special role in the theory. Natural
\textit{linear} isomorphisms between these spaces and $L^{2}\left(  E\right)
$ and its dual, respectively, will be exploited in \S 3.1.

\subsection{Bosonic Quantization of Admissible Quadratic Lagrangians}
We work here with a generalization of the fields in \S 1.1.  

\bigskip
\noindent {\bf Assumptions:}
\begin{enumerate}
\item  Replace the space $L^{2}\left(  E\right)  $ by a separable complex
Hilbert space $\mathcal{E}$, called the {\bf{space of classical fields}}.
The canonical pairing between $\mathcal{E}$ and its dual
$\mathcal{E}^{\ast}$ is denoted by $\left(  \cdot,\cdot\right)  :\mathcal{E}^{\ast}\times\mathcal{E}\rightarrow\mathbb{C}$.

\item {\label{gsen}} Replace the operator $\sqrt{m^{2}+\triangle_{E}}$ by a positive self-adjoint
{\bf{classical frequency operator}} ${\Omega:\mathcal{E}\rightarrow\mathcal{E}}$ which is bounded below 
by a positive constant $\lambda>0$. The {\bf{ground state energy}} ${\mu>0}$ is the
largest such $\lambda$.

\item  The compactness of $M$ is replaced by the assumption that $\Omega$ is
$\mathbf{\Theta}$\textbf{-summable}, i.e. that $\operatorname*{Tr}e^{-\beta\Omega}<\infty$ for all $\beta>0$.

\item  The \textbf{free Lagrangian} $\;\frak{L}:\mathcal{D}\left(  \Omega\right)  \times\Omega\rightarrow\mathbb{R}$ is given by
\begin{equation}
\frak{L}\left(  \varphi,d\varphi/dt\right)  =\left\langle d\varphi
/dt,d\varphi/dt\right\rangle -\left\langle \Omega\varphi,\Omega\varphi
\right\rangle .\label{abfreelag}\end{equation}
\end{enumerate}

\noindent Note that condition $\left(  \ref{gsen}\right)  $ rules out the case of a
massless $\left(  m=0\right)  $ scalar field on the circle $S^{1}$
parametrized by $\theta\in\left[  0,2\pi\right)  $ unless the Laplacian
$\triangle_{S^{1}}=-d^{2}/d\theta^{2}$ is \textit{twisted}. For $\rho$ not a
multiple of $2\pi$, the \textit{twisted Laplacian} $\triangle_{S^{1}}^{\rho}$
is the self-adjoint extension of $-d^{2}/d\theta^{2}$ acting on smooth
functions on $S^{1}$ which satisfy\[
\lim_{\theta\rightarrow2\pi^{-}}\frac{d^{n}f}{d\theta^{n}}=e^{i\rho}\frac{d^{n}f}{d\theta^{n}}\]
for $n\in\mathbb{N}$.

\begin{definition}
\label{defadmis2}\noindent An $\Omega$ satisfying the strict positivity and $\Theta$-summability
assumptions (2-3) is called {\bf{admissible}}, as is its associated free Lagrangian. 
The \textbf{canonical antilinear isomorphism }$f\mapsto
\bar{f}:\mathcal{E}\rightarrow\mathcal{E}^{\ast}$ is given by
\[
\left(  \bar{f},g\right)  =\left\langle f,g\right\rangle
\]
for all $g\in\mathcal{E}$. Given a linear or antilinear operator
$A:\mathcal{E}\rightarrow\mathcal{E}$, the \textbf{conjugate transformation}
$\bar{A}:\mathcal{E}^{\ast}\rightarrow\mathcal{E}^{\ast}$ is given by
\[
\bar{A}\bar{g}=\overline{Ag}.
\]
\end{definition}

To quantize this theory in the usual way, let $\{e_k\}_{k\in K}$ 
denote an
orthonormal basis of $\mathcal{E}$ consisting of eigenvectors of $\Omega$, 
namely, 
\begin{equation}
\Omega e_k = \omega_k e_k\ ,\quad {\rm for\ all\ }k\in K\ .
\end{equation}
Let $Op\left(  \frak{F}\right)  $ denote the set of linear operators on
$\frak{F}{}$. 
The \emph{real-time} quantum field $\varphi_{RT}:\mathbb{R}\times
\mathcal{E}^{\ast}\rightarrow Op\left(  \frak{F}\right)  $ is the
operator-valued function defined by\[
\varphi_{RT}\left(  t,\bar{f}\,\right)  =\sum_{k\in K}\frac{1}{\sqrt
{2\omega_{k}}}\left(  \alpha_{+}^{\ast}\left(  k\right)  e^{i\omega_{k}t}+\alpha_{-}\left(  -k\right)  e^{-i\omega_{k}t}\right)  \left(  \bar
{f},e_{k}\right)  \text{.}\]
The operators $\alpha_{\pm}\left(  \pm k\right)  $ are required to satisfy the
canonical commutation relations $\left(  \ref{CCR}\right)  $, and
to act on Fock space $\frak{F}$, which is isomorphic to the symmetric tensor
algebra $\exp_{\otimes_{S}}\mathcal{E}\oplus\mathcal{E}^{\ast}$.\footnote{We use $\alpha$ instead of the more standard variable $a$ to
remind the reader that the symbol $k$ need not correspond to momentum. The
notation $\alpha_{-}\left(  -k\right)  $ is preferred over $\alpha_{-}\left(
k\right)  $ to resemble the standard quantization of the free complex
scalar field in a rectangular box. In that case, the basis of $\mathcal{E}$
can be chosen to be of the form $e_{k}=e^{-ikx}$, where $k$ ranges over some
lattice. Then 
$\left(  \bar{e}_{k},f\right)  =\int e_{-k}\left(  x\right)  f\left(
x\right)  \,dx$
and $\bar{\alpha}_{-}\left(  -k\right)
=a_{-}\left(  -k\right)  $.}
The \emph{Hamiltonian} $H$ and \emph{particle number operator} $N$ of the system are defined as
\begin{align*}
H  & =\sum_{k\in K}\omega_k
\left(\alpha^*_+(k)\alpha_+(k)+\alpha^*_-(-k)\alpha_-(-k)\right)\ 
\\
N  & =\sum_{k\in K}\left(  \alpha_{+}^{\ast}\left(  k\right)  \alpha
_{+}\left(  k\right)  +\alpha_{-}^{\ast}\left(  -k\right)  \alpha_{-}\left(
-k\right)  \right)  \text{.}\end{align*}
The fields $\varphi_{RT}\left(  t,\bar{f}\,\right)  $ and
$\varphi_{RT}^{\ast}\left(  t,f\right)  $ commute and
$\varphi_{RT}\left(  t,\bar{f}\,\right)  \left(  N+1\right)
^{-1/2}$ is bounded. We then infer that the closure of 
$\varphi_{RT}\left(  t,\bar{f}\,\right)$ defined on the domain 
$\frak{F}_{0}$ that is algebraic subspace spanned by 
states of finite particle number, is normal.  This follows by an application
of Nelson's analytic vector theorem, Lemma 5.1 of  \cite{Nelson}, since the 
vectors in $\frak{F}_0$ are a common set of analytic vectors for the real and 
imaginary parts of the field.  
Note that for $\bar{f}\in\mathcal{D}\left(  \bar{\Omega}^{2}\right)  $, 
the field $\varphi_{RT}$ satisfies the Klein-Gordon equation\begin{equation}
\frac{\partial^{2}}{\partial t^{2}}\varphi_{RT}\left(
t,\bar{f}\right)  +\varphi_{RT}\left(  t,\bar{\Omega}^{2}\bar{f}\right)  =0,\label{kgprop}\end{equation}
where the derivative is taken strongly on $\frak{F}_{0}$.

The conjugate field $\varphi_{RT}^{\ast}:\mathbb{R}\times\mathcal{E}\rightarrow Op\left(  \frak{F}\right)  $ is given by\[
\varphi_{RT}^{\ast}\left(  t,f\right)  =\left(  \varphi_{RT}\left(  t,\bar
{f}\,\right)  \right)  ^{\ast}=\sum_{k\in K}\frac{1}{\sqrt{2\omega_{k}}}\left(  \alpha_{-}^{\ast}\left(  -k\right)  e^{i\omega_{k}t}+\alpha
_{+}\left(  k\right)  e^{-i\omega_{k}t}\right)  \left(  \bar{e}_{k},f\right)
\]
The \emph{imaginary-time} fields $\varphi:\left[  0,\infty\right)  \times\mathcal{E}^{\ast
}\rightarrow Op\left(  \frak{F}\right)  $ and $\bar{\varphi}:\left[  0,\infty\right)  \times\mathcal{E}\rightarrow Op\left(  \frak{F}\right)  $ are given by\[\begin{array}
[c]{ccc}\varphi\left(  t,\bar{f}\,\right)  =e^{-tH}\varphi_{RT}\left(  0,\bar
{f}\,\right)  e^{tH} & \text{,} & \bar{\varphi}\left(  t,f\right)
=e^{-tH}\varphi_{RT}^{\ast}\left(  0,f\right)  e^{tH}\end{array}
\]
Again, $\varphi\left(  t,\bar{f}\,\right)  $ and $\bar{\varphi}\left(
t,f\right)  $ give well-defined normal operators with core $\frak{\frak{F}}_{0}$ when $t{\geq}0$.

\subsection{Twist Positivity and the Twisted Pair Correlation Function}
The partition function of the system is defined as 
\begin{equation}
Z_\beta={\rm Tr}\left(e^{-\beta H}\right)\ ,
\end{equation}
so one needs the heat operator to be trace class. This is a 
consequence of the admissibility of $\Omega$.
\begin{definition}
For a unitary operator $U:\frak{F}\rightarrow\frak{F}$ commuting with the Hamiltonian $H$, the
\textbf{partition function twisted by U} is \[
Z_{\beta,U}=\operatorname*{Tr}\left(  Ue^{-\beta H}\right)  \text{.}\]
We say that $U$ is a \textbf{twist positive} with respect to $H$ if\[
Z_{\beta,U}>0
\]
for all $\beta>0$.
\end{definition}
It was observed in \cite{TwistPositivity} that, surprisingly enough, many
interesting symmetries $U$ are twist positivity, both in particle quantum
theory and in quantum field theory.
In this work, we generalize previous results by considering the following 
natural class of symmetries:
\begin{definition}
A bounded linear or antilinear operator $S:\mathcal{E}\rightarrow\mathcal{E} $
is a \textbf{Lagrangian\ symmetry}, if $S$ restricts to a bijection of
$\frak{D}\left(  \Omega\right)  $ onto itself and
\begin{equation}
\mathcal{L}\left(  \varphi,\dot{\varphi}\right)  =\mathcal{L}\left(
S\varphi,S\dot{\varphi}\right) \label{symdef1}\ .
\end{equation}
\end{definition}
Hence $S$ is a Lagrangian symmetry iff it preserves the closed quadratic
forms
\[\begin{array}
[c]{ccc}\varphi\mapsto\mathcal{L}\left(  \varphi,0\right)  & \text{and} &
d\varphi/dt\mapsto\mathcal{L}\left(  0,d\varphi/dt\right)
\end{array}
\]
and their domains. F{}rom the one-to-one correspondence between closed positive
quadratic forms and positive self-adjoint operators (see \cite{reedandsimonI}), it follows that $S$\textit{\ is a Lagrangian symmetry of
}$\mathcal{L}$\textit{\ iff }$\left[  S,\Omega\right]  =0$\textit{\ and }$S$\textit{\ is unitary or anitunitary.}

For each Lagrangian symmetry $S$, we shall denote by $U_S$ its
implementation on Fock space ${\frak{F}}$ (see \S 3).
The $U_S$ have a characteristic action on ${\frak{F}}$.
Indeed, the set of such
implementations is precisely the set of \textit{unitary\footnote{Note that an
antiunitary Lagrangian symmetry will have a unitary implementation on
$\frak{F}$.}} operators $U:\frak{F}\rightarrow\frak{F}$ such that

\begin{enumerate}
\item $U$ commutes with the Hamiltonian $H$, the number operator $N$, and with the combined time-charge
reversal operator $TC$, which is constructed in Theorem \ref{TCtheo} below.

\item $U$ preserves the vacuum, and $U$ acts independently on each particle in a multiparticle state, 
unaffected by the presense of other particles.

\item $U$ either sends particles to particles of the same charge (for $S$ unitary) or to particles of opposite charge (for $S$ antiunitary).
\end{enumerate}
That properties 1-3 hold for implementations of Lagrangian symmetries is proven in lemmas  \ref{goodprops} and \ref{ctlemma}. Theorem \ref{applyiso} implies that all such symmetries $U$ are implementations of Lagrangian symmetries.

We now give our main results.  

\noindent
\begin{theorem}
The Fock space implementation $U_S$ of a
Lagrangian symmetry $S$ of an admissible free Lagrangian
is twist positive.
\end{theorem}

We define the twisted pair correlation function and associated objects:
\begin{definition}
\label{defcintro}
Let $0\leq t,s\leq\beta$, $\bar{f}\in \mathcal{E}^*$ and $g\in \mathcal{E}$.
The {\bf{time-ordered product}} is given by
\begin{equation}
\label{timeorder}
(\varphi(t,\bar{f})\bar{\varphi}(s,g))_+=
\left\{ \begin{array}{ll}
\varphi(t,\bar{f})\bar{\varphi}(s,g) & {\rm if\ } t<s\\
\bar{\varphi}(s,g)\varphi(t,\bar{f}) & {\rm if\ } t\geq s
\end{array}
\right.
\end{equation}
For a unitary Lagrangian symmetry $S$, 
the {\bf{twisted pair correlation function}} {$\mathbf{C_{\beta,U_S}}$} is 
\begin{equation}
C_{\beta,U_S}(t,\bar{f};s,g)=\frac{1}{Z_{\beta,U_S}}
{\rm Tr}\left((\varphi(t,\bar{f})\bar{\varphi}(s,g))_+
U_Se^{-\beta H}\right)\ .
\end{equation}
The twisted pair correlation function is the integral kernel of the
{\bf{twisted pair correlation operator}} ${\mathbf{C_{\beta}}}$ on the {\bf{path
space}} $\mathcal{T}_{\beta}=L^{2}\left(  \left[  0,\beta\right)
;\mathcal{E}\right)  \cong L^{2}\left(  \left[  0,\beta\right)
\right)  \otimes\mathcal{E}$ of square-integrable functions from $\left[
0,\beta\right)  $ to $\mathcal{E}$.
The operator $C_{\beta}$ is that which satisfies\begin{equation}
\left\langle \tilde{f},C_{\beta}\tilde{g}\right\rangle _{\mathcal{T}_{\beta}}=\int_{0}^{\beta}\int_{0}^{\beta}C_{\beta,U_S}\left(  t,\overline{\tilde{f}\left(
t\right)  };s,\tilde{g}\left(  t\right)  \right)  \;ds\,dt ,
\end{equation}
where $\tilde{f},\tilde{g}\in\mathcal{T}_{\beta}$.
Define the {\bf{twisted derivative}} $\mathbf{D}$ as $i$ times the self-adjoint extension of 
$\frac{1}{i}\frac{\partial}{\partial t}$ acting on smooth functions
$\widetilde{g}:[0,\beta)\rightarrow \mathcal{E}$ such that for all 
$n\in{\mathbb N}$,
\begin{equation}
\frac{\partial^{n}\tilde{g}}{\partial t^{n}}\left(  0\right)  =\lim\limits_{t\rightarrow\beta^{-}}S^{\ast}\frac{\partial^{n}\tilde{g}}{\partial t^{n}}\left(  t\right)
\ .\label{funboundry}\end{equation}
\end {definition}
We prove in \S 5 the following\newline

\noindent
\begin{theorem}
If $S$ is a unitary Lagrangian symmetry of an admissible free Lagrangian
then 
\begin{equation}
C_{\beta}=\left(  -D^{2}+\Omega^{2}\right)  ^{-1},\end{equation}
where $\Omega$ is identified with $I\otimes\Omega$.
\end{theorem}
In \S 6, we modify and extend this theorem to antiunitary symmetries
and to the case of real scalar fields.

The positivity of the operator $C_\beta$ ensures the existence of a 
countably additive Borel measure whose moments are the twisted correlation 
functions of the free field.  It would be interesting to extend the
techniques of constructive quantum field theory to non-linear
perturbations of the free theories we consider here.

\pagebreak

\section{\smallskip Bosonic Quantization of Complex Free Fields}

\subsection{The standard $L^{2}\left(  X\right)  $ representation of
$\mathcal{E}$.\label{basismethod}}

Making contact with the standard physics notation for complex free Bosonic
fields, we represent the space $\mathcal{E}$ as $L^{2}\left(  X\right)  $, for
some measure space $X$, so that expressions involving fields $\varphi
\in\mathcal{E}$ may be written in a familiar form as $\varphi\left(  x\right)
$, $x\in X$. We note that if one identifies a function $f\in L^{2}\left(
X\right)  $ with the linear functional on $L^{2}\left(  X\right)  $ given by
\[
g\mapsto\int f\left(  x\right)  g\left(  x\right)  \,dx\,\text{,}%
\]
then the canonical isomorphism $\bar{\cdot}$ is just complex conjugation, and
the conjugate transformation of $A$ is given by
\[
\bar{A}f=\overline{\left(  A\bar{f}\,\right)  }\text{,}%
\]
where the bars on the right-side denote complex conjugation. Elements of
$\mathcal{E}^{\ast}$ will always be represented below as $\bar{f}$, for some
element $f\in\mathcal{E}$. In particular, we make no essential use of complex
conjugation on $L^{2}\left(  X\right)  $, which is not a natural
representation-independent operation on $\mathcal{E}$.\footnote{An example of
a complex classical field theory given on an $L^{2}$ space with physically
unnatural conjugation is the complex scalar field on the ``twisted circle.''
Set $\mathcal{E}=L^{2}\left(  S^{1}\right)  $, $\Omega=\triangle_{S^{1}}%
^{\rho}$, where $\triangle_{S^{1}}^{\rho}$ is the twisted Laplacian defined in
\S\ref{firstsect11}. Hence unless $\rho$ is a multiple of $\pi$, we see that
the $L^{2}$-conjugation on $\mathcal{E}$ is does not commute with least-action
time evolution.}

The operator-valued linear functionals $\mathbf{\varphi}_{RT} $
and $\mathbf{\varphi}_{RT}^{\ast}$ are commonly expressed in
suggestive notation as
\begin{equation}
\varphi_{RT}\left(  t,\bar{f}\,\right)  =\int\varphi
_{RT}\left(  t,x\right)  \bar{f}\left(  x\right)
dx\label{functional0}%
\end{equation}
and
\begin{equation}
\varphi_{RT}^{\ast}\left(  t,f\right)  =\int\varphi
_{RT}^{\ast}\left(  t,x\right)  f\left(  x\right)
dx\text{.\label{functional}}%
\end{equation}
Here $\varphi_{RT}\left(  t,x\right)  $ and $\varphi
_{RT}^{\ast}\left(  t,x\right)  $ are notational devices
expressed as
\begin{align*}
\varphi_{RT}\left(  t,x\right)   &  =\sum_{k}\frac{1}%
{\sqrt{2\omega_{k}}}\left(  e^{i\omega_{k}t}\alpha_{+}^{\ast}\left(  k\right)
+e^{-i\omega_{k}t}\alpha_{-}\left(  -k\right)  \right)  e_{k}\left(  x\right)
\\
\varphi_{RT}^{\ast}\left(  t,x\right)   &  =\sum_{k}\frac
{1}{\sqrt{2\omega_{k}}}\left(  e^{i\omega_{k}t}\alpha_{-}^{\ast}\left(
-k\right)  +e^{-i\omega_{k}t}\alpha_{+}\left(  k\right)  \right)  \bar{e}%
_{k}\left(  x\right)  \text{,}%
\end{align*}
where the summations are understood to be interchanged with the integrals in
$\left(  \ref{functional0}-\ref{functional}\right)  $. Note that
$\varphi_{RT}\left(  t,x\right)  $ and $\varphi
_{RT}^{\ast}\left(  t,x\right)  $ are not functions, and
pointwise they are merely symbolic expressions.

As usual, the canonical commutation relations for the operators $\alpha_{\pm}$
are equivalent to the equal-time canonical commutators:
\begin{align}
\left[  \varphi_{RT}\left(  t,\bar{f}\,\right)  ,\frac
{\partial\varphi_{RT}^{\ast}}{\partial t}\left(  t,g\right)
\right]   &  =i\left(  \bar{f},g\right)  =i\int\bar{f}\left(  x\right)
g\left(  x\right)  dx=i\left\langle f,g\right\rangle \label{ccr1}\\
\left[  \varphi_{RT}\left(  t,\bar{f}\,\right)  ,\frac
{\partial\varphi_{RT}}{\partial t}\left(  t,\bar{g}\right)
\right]   &  =\left[  \varphi_{RT}\left(  t,\bar{f}\,\right)
,\varphi_{RT}^{\ast}\left(  t,g\right)  \right]  =\left[
\varphi_{RT}\left(  t,\bar{f}\,\right)  ,\varphi
_{RT}\left(  t,\bar{g}\,\right)  \right]  =0\text{.}%
\label{ccr2}%
\end{align}
Here $\partial\varphi_{RT}/\partial t$ and $\partial
\varphi_{RT}^{\ast}/\partial t$ are defined using the strong
limit on $\frak{F}_{0}$.

The Klein-Gordon equation $\left(  \ref{kgprop}\right)  $ for $\varphi
_{\text{cl}}\left(  t,x\right)  $ is denoted by
\begin{equation}
\left(  \partial_{t}^{2}+\Omega_{x}^{2}\right)  \varphi_{\text{cl}}\left(
t,x\right)  =0\text{,}\label{classkg}%
\end{equation}
and the \textit{conjugate} Klein-Gordon equation for $\bar{\varphi}%
_{\text{cl}}\left(  t,x\right)  $ is
\begin{equation}
\left(  \partial_{t}^{2}+\bar{\Omega}_{x}^{2}\right)  \bar{\varphi}%
_{\text{cl}}\left(  t,x\right)  =0\text{.}%
\end{equation}
Notice that when $\mathcal{E}$ is the space $L^{2}\left(  E\right)  $ of
square integrable sections of the vector bundle $E$ then $\bar{\varphi
}_{\text{cl}}$ is a section of the dual bundle $E^{\ast}$. This is why we
refrained from identifying the space $\mathcal{E}$ with its dual.

\subsection{Creation/Annihilation Functionals \& Basis-Free
Quantization\label{nobasissec}}

We introduce the creation and annihilation functionals, which play a role in
the Fock space implementation of classical symmetries. We show that they are
natural objects which come from a basis-independent method of quantization.

\begin{definition}
The \textbf{linear creation functionals} $A_{+}^{\ast}:\mathcal{E}^{\ast
}\rightarrow Op\left(  \frak{F}\right)  $ and $A_{-}^{\ast}:\mathcal{E}%
\rightarrow Op\left(  \frak{F}\right)  $ are defined by
\begin{align*}
A_{+}^{\ast}\left(  \bar{f}\,\right)   &  =\sum_{k}\left(  \bar{f}%
,e_{k}\right)  \alpha_{+}^{\ast}\left(  k\right)  =\sum_{k}\left\langle
f,e_{k}\right\rangle _{\mathcal{E}}\alpha_{+}^{\ast}\left(  k\right)  \text{,
\ \ \ \ }\bar{f}\in\mathcal{E}^{\ast}\\
A_{-}^{\ast}\left(  f\right)   &  =\sum_{k}\left(  \bar{e}_{k},f\right)
\alpha_{-}^{\ast}\left(  -k\right)  =\sum_{k}\left\langle e_{k},f\right\rangle
_{\mathcal{E}}\alpha_{-}^{\ast}\left(  -k\right)  \text{,\ \ }f\in
\mathcal{E}\text{.}%
\end{align*}
The linear creation functionals are well-defined operators on $\frak{F}_{0}$.
The \textbf{linear} \textbf{annihilation functionals}\textit{\ }are given by
\begin{align*}
A_{+}\left(  f\right)   &  =\left(  A_{+}^{\ast}\left(  \bar{f}\,\right)
\right)  ^{\ast}=\sum_{k}\left(  \bar{e}_{k},f\right)  \alpha_{+}\left(
k\right)  =\sum_{k}\left\langle e_{k},f\right\rangle _{\mathcal{E}}\alpha
_{+}\left(  k\right) \\
A_{-}\left(  \bar{f}\,\right)   &  =\left(  A_{-}^{\ast}\left(  f\right)
\right)  ^{\ast}=\sum_{k}\left(  \bar{f},e_{k}\right)  \alpha_{-}\left(
-k\right)  =\sum_{k}\left\langle f,e_{k}\right\rangle _{\mathcal{E}}\alpha
_{-}\left(  -k\right)  .
\end{align*}
\end{definition}

We state without proof the following

\begin{theorem}
\label{gencre}The creation and annihilation functionals satisfy for all
$f,g\in\mathcal{E}$
\begin{align}
\left[  A_{+}\left(  f\,\right)  ,A_{+}^{\ast}\left(  \bar{g}\right)
\right]   &  =\left(  \bar{g},f\right)  =\left\langle g,f\right\rangle
_{\mathcal{E}}\label{genccr1}\\
\left[  A_{-}\left(  \bar{f}\,\right)  ,A_{-}^{\ast}\left(  g\right)
\right]   &  =\left(  \bar{f},g\right)  =\left\langle f,g\right\rangle
_{\mathcal{E}}\text{\thinspace.}\label{genccr2}%
\end{align}
The dynamics of the $A_{\pm}^{\ast}$ are given by
\begin{align}
e^{itH}A_{+}^{\ast}\left(  \bar{f}\,\right)  e^{-itH} &  =A_{+}^{\ast}\left(
e^{it\bar{\Omega}}\bar{f}\right) \label{gdinam1}\\
e^{itH}A_{-}^{\ast}\left(  f\right)  e^{-itH} &  =A_{-}^{\ast}\left(
e^{it\Omega}f\right)  \,.\label{gdinam2}%
\end{align}
Furthermore\footnote{For $\mu>0$, equations $\left(  \ref{defphi2}\right)
-\left(  \ref{defphi23}\right)  $ show that if we restrict $\varphi
_{RT}\left(  t,\bar{f}\,\right)  $ and $\varphi
_{RT}^{\ast}\left(  t,f\right)  $ to the subspace of $\frak{F}$
containing states of at most $n$ particles, for fixed $n<\infty$, then we may
continuously extend $\varphi_{RT}\left(  t,\bar{f}\,\right)  $
and $\varphi_{RT}^{\ast}\left(  t,f\right)  $ to $f\in
\mathcal{E}^{-1/2}$, the completion of $\mathcal{E}$ in the inner product
$\left\langle f,g\right\rangle _{-1/2}=\left\langle \Omega^{-1/2}%
f,\Omega^{-1/2}g\right\rangle $.}
\begin{align}
\varphi_{RT}\left(  t,\bar{f}\,\right)   &  =\frac{1}{\sqrt{2}%
}\left[  A_{+}^{\ast}\left(  \bar{\Omega}^{-1/2}e^{it\bar{\Omega}}\bar
{f}\,\right)  +A_{-}\left(  \bar{\Omega}^{-1/2}e^{-it\bar{\Omega}}\bar
{f}\,\right)  \right] \label{defphi2}\\
\varphi_{RT}^{\ast}\left(  t,f\right)   &  =\frac{1}{\sqrt{2}%
}\left[  A_{-}^{\ast}\left(  \Omega^{-1/2}e^{it\Omega}f\right)  +A_{+}\left(
\Omega^{-1/2}e^{-it\Omega}f\right)  \right]  \text{,}\label{defphi23}%
\end{align}
and for $f\in\mathcal{D}\left(  e^{t\Omega}\right)  $
\begin{align}
\varphi\left(  t,\bar{f}\,\right)   &  =\frac{1}{\sqrt{2}}\left[  A_{+}^{\ast
}\left(  \bar{\Omega}^{-1/2}e^{-t\bar{\Omega}}\bar{f}\,\right)  +A_{-}\left(
\bar{\Omega}^{-1/2}e^{t\bar{\Omega}}\bar{f}\,\right)  \right] \label{pheima}\\
\bar{\varphi}\left(  t,f\right)   &  =\frac{1}{\sqrt{2}}\left[  A_{-}^{\ast
}\left(  \Omega^{-1/2}e^{-t\Omega}f\right)  +A_{+}\left(  \Omega
^{-1/2}e^{t\Omega}f\,\right)  \right]  \,.\label{pheimb}%
\end{align}
If we define the maps
\begin{align*}
\Gamma_{+} &  :\mathcal{E}^{\ast}\rightarrow\frak{F},\;\;\text{ }\bar
{f}\mapsto A_{+}^{\ast}\left(  \bar{f}\,\right)  \left|  0\right\rangle \\
\Gamma_{-} &  :\mathcal{E}^{\;}\rightarrow\frak{F},\text{ \ \ }f\mapsto
A_{-}^{\ast}\left(  f\right)  \left|  0\right\rangle
\end{align*}
then each $\Gamma_{\pm}$ is a linear Hilbert space isomorphism onto the
appropriately charged $1$-particle subspace of $\frak{F}$,\footnote{Note that
to any given linear Hilbert-space isomorphism $\tilde{\Gamma}_{+}%
^{\;}:\mathcal{E}\rightarrow\frak{F}_{+}^{\left(  1\right)  }%
=\operatorname*{Span}\left\{  \alpha_{+}^{\ast}\left(  k\right)  \left|
0\right\rangle \right\}  $ corresponds the antiunitary operator $f\mapsto
\tilde{\Gamma}_{+}^{\ast}\Gamma_{+}\left(  \bar{f}\text{\thinspace}\right)  $.
Hence \emph{a natural linear isomorphism between} $\mathcal{E}$ \emph{and
}$\frak{F}_{1}^{+}$ \emph{exists only if }$\mathcal{E}$ \emph{is equipped with
a preferred antiunitary operator.} Such is the case neither if $\mathcal{E}$
is produced by the Stone-von Neumann theorem from the (exponentiated)
quantum-mechanical canonical commutation relations nor if $\mathcal{E}$ is the
set of square-integrable sections of an arbitrary vector-bundle.} and
\begin{align*}
\Gamma_{+}^{\ast}H\Gamma_{+} &  =\bar{\Omega}\\
\Gamma_{-}^{\ast}H\Gamma_{-} &  =\Omega\,.
\end{align*}
\end{theorem}

\smallskip As promised, we now sketch an equivalent quantization which does
not rely on a arbitrary choice of basis. Given a solution $\varphi_{\text{cl}%
}$ to the classical equations of motion $\left(  \ref{classkg}\right)  $,
define the (unquantized) linear functionals $\bar{A}_{+}$ and $\bar{A}_{-}$ on
$\mathcal{E}^{\ast}$ and $\mathcal{E}$, respectively, by setting
\begin{align}
\bar{A}_{+}\left(  \bar{f}\,\right)   &  =\frac{1}{\sqrt{2}}\int
\varphi_{\text{cl}}\left(  t,x\right)  \left(  \bar{\Omega}^{1/2}%
e^{-it\bar{\Omega}}\bar{f}\right)  \left(  x\right)  -i\frac{\partial
\varphi_{\text{cl}}\left(  t,x\right)  }{\partial t}\left(  \bar{\Omega
}^{-1/2}e^{-i\bar{\Omega}t}\bar{f}\,\right)  \left(  x\right)
\;dx\label{ggeq1}\\
\bar{A}_{-}\left(  f\right)   &  =\frac{1}{\sqrt{2}}\int\bar{\varphi
}_{\text{cl}}\left(  t,x\right)  \left(  \Omega^{1/2}e^{-it\Omega}f\right)
\left(  x\right)  -i\frac{\partial\bar{\varphi}_{\text{cl}}\left(  t,x\right)
}{\partial t}\left(  \Omega^{-1/2}e^{-i\Omega t}f\right)  \left(  x\right)
\;dx\label{geq2}%
\end{align}
for $f\in\mathcal{E}$. Equations $\left(  \ref{ggeq1}\right)  -\left(
\ref{geq2}\right)  $ are independent of $t,$ since $\varphi_{\text{cl}}$
satisfies the Klein-Gordon equation, eq$.$ $\left(  \ref{classkg}\right)  $.
We then replace the $\bar{A}_{\pm}$ by operator-valued linear functionals
$A_{\pm}^{\ast}$ satisfying $\left(  \ref{genccr1}-\ref{genccr2}\right)  $.
The quantized field $\varphi_{RT}\left(  t,\bar{f}\,\right)  $
is then defined by equation $\left(  \ref{defphi2}\right)  $. The Fock space
is defined in the obvious way, and the Hamiltonian may be defined by
\[
H=\sum_{k}A_{+}^{\ast}\left(  \bar{\Omega}^{1/2}\bar{f}_{k}\right)
A_{+}\left(  \Omega^{1/2}f_{k}\right)  +\sum_{k}A_{-}^{\ast}\left(
\Omega^{1/2}f_{k}\right)  A_{-}\left(  \bar{\Omega}_{k}^{1/2}\bar{f}%
_{k}\right)  \text{,}%
\]
where $\left\{  f_{k}\right\}  $ is an arbitrary orthonormal basis of
$\mathcal{E}$. The quantization of \S1.2 may be recovered using the equations
$\alpha_{+}^{\ast}\left(  k\right)  =A_{+}^{\ast}\left(  \bar{e}_{k}\right)
$, $\alpha_{-}^{\ast}\left(  -k\right)  =A_{-}^{\ast}\left(  e_{k}\right)  $.

\section{\smallskip Implementing Lagrangian Symmetries on Fock Space}

To motivate the definition of the Fock-space implementation of Lagrangian
symmetries, we examine the adjoint substitution of test-functions which
implements a unitary Lagrangian symmetry $S:\mathcal{E}\rightarrow\mathcal{E}$
at the classical level. If we replace $\varphi_{\text{cl}}\rightarrow
S\varphi_{\text{cl}}$ then
\begin{equation}
\int\varphi_{\text{cl}}\left(  t,x\right)  \bar{f}\left(  x\right)
dx\rightarrow\int S\varphi_{\text{cl}}\left(  t,x\right)  \bar{f}\left(
x\right)  dx=\left\langle f\left(  \cdot\right)  ,S\varphi_{\text{cl}}\left(
t,\cdot\right)  \right\rangle _{\mathcal{E}}=\int\varphi_{\text{cl}}\left(
t,x\right)  \left(  \bar{S}^{\ast}\bar{f}\,\right)  \left(  x\right)
dx\label{phigoto}%
\end{equation}
where $\bar{S}^{\ast}\equiv\left(  \bar{S}\right)  ^{\ast}=\overline{\,\left(
S^{\ast}\right)  \,}$. Similarly,
\begin{equation}
\int\bar{\varphi}_{\text{cl}}\left(  t,x\right)  f\left(  x\right)
dx\rightarrow\int\left(  \,\overline{S\varphi}_{\text{cl}}\right)  \left(
t,x\right)  f\left(  x\right)  dx=\int\bar{\varphi}_{\text{cl}}\left(
t,x\right)  \left(  S^{\ast}f\right)  \left(  x\right)  dx\text{.}%
\label{phibgoto}%
\end{equation}
The first and second transformations are implementable by the substitutions
$\bar{f}\rightarrow\bar{S}^{\ast}\bar{f}$ and $f\rightarrow S^{\ast}f$,
respectively. We use these adjoint substitutions (and similar considerations
for antiunitary symmetries) as our definition:

\begin{definition}
\label{Usdef}Let $S:\mathcal{E}\rightarrow\mathcal{E}$ be a Lagrangian
symmetry. For $S$ unitary, the corresponding \textbf{Fock-space
implementation} $\mathbf{U}_{S}:\frak{F}\rightarrow\frak{F}$ is the linear
operator which satisfies
\begin{align}
U_{S}\left|  0\right\rangle  &  =\left|  0\right\rangle \nonumber\\
U_{S}\varphi_{RT}\left(  t,\bar{f}\,\right)  U_{S}^{\ast} &
=\varphi_{RT}\left(  t,\bar{S}^{\ast}\bar{f}\,\right)
\label{smoothdef1}\\
U_{S}\varphi_{RT}^{\ast}\left(  t,f\right)  U_{S}^{\ast} &
=\varphi_{RT}^{\ast}\left(  t,S^{\ast}f\right)  \text{.}%
\label{smoothdef2}%
\end{align}
If $S$ is an antiunitary symmetry then $\mathbf{U}_{S}:\frak{F}\rightarrow
\frak{F}$ is given by
\begin{align*}
U_{S}\left|  0\right\rangle  &  =\left|  0\right\rangle \\
U_{S}\varphi_{RT}\left(  t,\bar{f}\,\right)  U_{S}^{\ast} &
=\varphi_{RT}^{\ast}\left(  t,S^{\ast}f\right) \\
U_{S}\varphi_{RT}^{\ast}\left(  t,f\right)  U_{S}^{\ast} &
=\varphi_{RT}\left(  t,\bar{S}^{\ast}\bar{f}\,\right)  \text{.}%
\end{align*}
Note that for antiunitary $S$ the adjoint $S^{\ast}$ satisfies
\[
\left\langle f,Sg\right\rangle =\overline{\left\langle S^{\ast}%
f,g\right\rangle }\text{.}%
\]
\end{definition}

The following lemma gives the properties of $U_{S}$:

\begin{lemma}
\label{goodprops}Let $S$ and $V$ be a unitary and an anti-unitary Lagrangian
symmetries, respectively. Then $U_{S}$ and $U_{V}$ exist, are unitary, and
commute with $H$. Furthermore, the actions of $U_{S}$ and $U_{V}$ on
$\frak{F}{}$ are given by
\begin{align}
U_{S}\left|  0\right\rangle  &  =\left|  0\right\rangle \nonumber\\
U_{S}A_{+}^{\ast}\left(  \bar{f}\,\right)  U_{S}^{\ast} &  =A_{+}^{\ast
}\left(  \bar{S}^{\ast}\bar{f}\,\right) \label{ickdef1}\\
U_{S}A_{-}^{\ast}\left(  f\,\right)  U_{S}^{\ast} &  =A_{-}^{\ast}\left(
S^{\ast}f\,\right) \label{ickdef2}%
\end{align}
and
\begin{align*}
U_{V}\left|  0\right\rangle  &  =\left|  0\right\rangle \\
U_{V}A_{+}^{\ast}\left(  \bar{f}\,\right)  U_{V}^{\ast} &  =A_{-}^{\ast
}\left(  V^{\ast}f\,\right) \\
U_{V}A_{-}^{\ast}\left(  f\,\right)  U_{V}^{\ast} &  =A_{+}^{\ast}\left(
\bar{V}^{\ast}\bar{f}\,\right)  \text{.}%
\end{align*}
\end{lemma}%

%TCIMACRO{\TeXButton{Proof}{\proof}}%
%BeginExpansion
\proof
%EndExpansion
If $U_{S}$ exists then it follows from $\left(  \ref{ggeq1}\right)  -\left(
\ref{geq2}\right)  $ that it satisfies $\left(  \ref{ickdef1}\right)
$-$\left(  \ref{ickdef2}\right)  $. Existence and unitarity follow from the
tensor product structure of $\frak{F}$. The fact that $U_{S}$ commutes with
$H$ follows from $\left(  \ref{gdinam1}\right)  -\left(  \ref{gdinam2}\right)
$.

We omit the similar proof of the antiunitary case.{\hfill$\blacksquare$}

\subsection{$TC$ Invariance}

\begin{theorem}
\label{TCtheo}There is a unique antilinear operator $TC$ on $\frak{F}$ such
that $TC\left|  0\right\rangle =\left|  0\right\rangle $ and
\begin{align*}
TC\,\varphi_{RT}\left(  t,\bar{f}\,\right)  \,\left(
TC\right)  ^{-1} &  =\varphi_{RT}^{\ast}\left(  -t,f\right) \\
TC\,\varphi_{RT}^{\ast}\left(  t,f\right)  \,\left(  TC\right)
^{-1} &  =\varphi_{RT}\left(  -t,\bar{f}\,\right)  \text{.}%
\end{align*}
Furthermore, $TC$ is antiunitary, squares to the identity, and satisfies
\begin{align*}
TC\,A_{+}^{\ast}\left(  \bar{f}\,\right)  \,\left(  TC\right)  ^{\ast} &
=A_{-}^{\ast}\left(  f\right) \\
TC\,A_{-}^{\ast}\left(  f\right)  \,\left(  TC\right)  ^{\ast} &  =A_{+}%
^{\ast}\left(  \bar{f}\,\right)  \text{.}%
\end{align*}
\end{theorem}

The proof is similar to that of lemma \ref{goodprops} and is omitted. Notice
that by the CCR $\left(  \ref{ccr1}\right)  -\left(  \ref{ccr2}\right)  $,
there are no linear or antilinear operators $\tilde{T}$ and $\tilde{C}$ on
$\frak{F}$ with the properties that
\[%
\begin{array}
[c]{lll}%
\tilde{T}\varphi_{RT}\left(  t,\bar{f}\,\right)  \tilde{T}%
^{-1}=\varphi_{RT}\left(  -t,\bar{f}\,\right)  & \text{,} &
\tilde{T}\varphi_{RT}^{\ast}\left(  t,f\right)  \tilde{T}%
^{-1}=\varphi_{RT}^{\ast}\left(  -t,f\,\right)
\end{array}
\]
and
\[%
\begin{array}
[c]{lll}%
\tilde{C}\varphi_{RT}\left(  t,\bar{f}\,\right)  \tilde{C}%
^{-1}=\varphi_{RT}^{\ast}\left(  t,f\right)  & \text{,} &
\tilde{C}\varphi_{RT}^{\ast}\left(  t,f\right)  \tilde{C}%
^{-1}=\varphi_{RT}\left(  t,\bar{f}\,\right)  \text{.}%
\end{array}
\]
However, if the Hilbert space $\mathcal{E}$ carries a conjugation\footnote{A
conjugation is an antiunitary map that squares to $1$.} $^{\vee}$ which
commutes with $\Omega$, then it is easy to verify that there is an antiunitary
operator $T^{\vee}$ and a unitary operator $C^{\vee}$ on $\frak{F}$ such that
\begin{align*}
T^{\vee}\left|  0\right\rangle  &  =C^{\vee}\left|  0\right\rangle =\left|
0\right\rangle \\
T^{\vee}\varphi_{RT}\left(  t,\bar{f}\,\right)  T^{\vee-1} &
=\varphi_{RT}\left(  -t,\overline{f^{\vee}}\,\right) \\
T^{\vee}\varphi_{RT}^{\ast}\left(  t,f\right)  T^{\vee-1} &
=\varphi_{RT}\left(  -t,f^{\vee}\right) \\
C^{\vee}\varphi_{RT}\left(  t,\bar{f}\,\right)  C^{\vee-1} &
=\varphi_{RT}^{\ast}\left(  t,f^{\vee}\right) \\
C^{\vee}\varphi_{RT}^{\ast}\left(  t,f\right)  C^{\vee-1} &
=\varphi_{RT}\left(  t,\overline{f^{\vee}}\,\right)  \text{.}%
\end{align*}
Furthermore, $T^{\vee}C^{\vee}=TC$.

$TC$ symmetry plays a crucial role in the proofs to follow. We summarize
behavior of $U_{S}$ and $\varphi$ under $TC$ symmetry:

\begin{lemma}
\label{ctlemma}Let $S$ $:\mathcal{E}\rightarrow\mathcal{E}$ be a Lagrangian
symmetry. Then

\begin{enumerate}
\item $\left[  U_{S},TC\right]  =0$

\item $TC\,\varphi\left(  t,\bar{f}\,\right)  \,TC=\bar{\varphi}\left(
t,f\right)  $

\item $TC\,\bar{\varphi}\left(  t,g\right)  \,TC=\varphi\left(  t,\bar
{g}\right)  $
\end{enumerate}
\end{lemma}

Finally, we state the following Theorem, which shows that the class of
Lagrangian symmetries contains most of the symmetries encountered in practice.

\begin{theorem}
\label{applyiso}Let $U$ be a unitary operator on the one-particle subspace
$\frak{F}^{\left(  1\right)  }$ of $\frak{F}$ such that

\begin{enumerate}
\item $U$ commutes with $H$ and $TC$.

\item $U$ maps $\frak{F}_{\pm}^{\left(  1\right)  }$ either to itself or to
$\frak{F}_{\mp}^{\left(  1\right)  }$.\footnote{$\frak{F}_{+}^{\left(
1\right)  }$ denotes the subspace of $\frak{F}^{\left(  1\right)  }$
consisting of elements of the form $A_{+}^{*}\left(  \bar{f}\right)  \left|
0\right\rangle $. $\frak{F}_{-}^{\left(  1\right)  }$ is defined analogously.}
\end{enumerate}

\noindent Then there exists a unique Lagrangian symmetry $S$ such that
\[
U=\left.  U_{S}\right|  _{\frak{F}^{\left(  1\right)  }}%
\]
\end{theorem}

The proof is a simple application of the isomorphisms $\Gamma_{\pm}$ of
Theorem \ref{gencre}.

\section{Twist Positivity}

Having finished with our investigation of the Fock-space implementations of
Lagrangian symmetries, we may now prove the anticipated theorems. That the
partition function is well-defined follows from

\begin{lemma}
\label{admisbound}If $\Omega$ is an admissible classical frequency operator
(see definition \ref{defadmis2}) then $e^{-\beta H}$ is trace-class.
\end{lemma}%

%TCIMACRO{\TeXButton{Proof}{\proof}}%
%BeginExpansion
\proof
%EndExpansion
Since $\omega_{k}>0$,
\[
\operatorname*{Tr}\left(  e^{-\beta H}\right)  =\prod_{k}\left(  \frac
{1}{1-e^{-\beta\omega_{k}}}\right)  ^{2}=\left(  \prod_{k}\left(
1+\frac{e^{-\beta\omega_{k}}}{1-e^{-\beta\omega_{k}}}\right)  \right)
^{2}\text{,}%
\]
the conclusion follows from the estimate
\begin{equation}
\sum\limits_{k}\frac{e^{-\beta\omega_{k}}}{1-e^{-\beta\omega_{k}}}\leq\frac
{1}{1-e^{-\beta\mu}}\operatorname*{Tr}\left(  e^{-\beta\Omega}\right)
\text{.}\tag*{$\blacksquare$}%
\end{equation}

\noindent The next theorem shows that many symmetry operators are twist positive.

\begin{theorem}
\label{tptheorem}Let $S:\mathcal{E}\rightarrow\mathcal{E}$ be a linear or
antilinear Lagrangian symmetry of an admissible free Lagrangian $\frak{L}$.
Then the Fock space implementation $U_{S}$ of $S$ is twist positive.
Furthermore, for antiunitary $S$ we have
\begin{equation}
\operatorname*{Tr}\left(  U_{S}e^{-\beta H}\right)  =\sqrt{\operatorname*{Tr}%
\left(  U_{S^{2}}e^{-2\beta H}\right)  }\text{.}\label{suggestive}%
\end{equation}
\end{theorem}

We note that twist \textit{nonnegativity }is a consequence $TC$ symmetry
(Lemma \ref{ctlemma}).%

%TCIMACRO{\TeXButton{Proof}{\proof}}%
%BeginExpansion
\proof
%EndExpansion
We first consider the case that $S$ is unitary. Choosing the basis $\left\{
e_{k}\right\}  $ of section \ref{basismethod} to simultaneously diagonalize
$\Omega$ and $S$,\footnote{Unitarity is used here, since an antiunitary
operator is diagonalizable only if it is a conjugation.}
\begin{align}
\Omega e_{k} &  =\omega_{k}e_{k}\label{ddd1}\\
S\,e_{k} &  =\rho_{k}e_{k}\label{ddd2}%
\end{align}
we compute
\begin{equation}
\operatorname*{Tr}\nolimits_{\frak{F}}\left(  U_{S}e^{-\beta H}\right)
=\prod_{k}\frac{1}{\left|  1-\rho_{k}e^{-\beta\omega_{k}}\right|  ^{2}%
}\text{.}\label{gottraceuebh}%
\end{equation}
Twist positivity follows, since
\[
\prod_{k}\frac{1}{\left|  1-\rho_{k}e^{-\beta\omega_{k}}\right|  ^{2}}%
\geq\left(  \prod_{k}\frac{1}{1+e^{-\beta\omega_{k}}}\right)  ^{2}%
=e^{-2\sum\limits_{k}\log\left(  1+e^{-\beta\omega_{k}}\right)  }\geq
e^{-2\operatorname*{Tr}\limits_{\mathcal{E}}e^{-\beta\Omega}}>0\text{.}%
\]

Although in section $\ref{realfieldsect}$ below we shall see that the previous
proof may be altered to include the antiunitary case,\footnote{One may also
use the (conjugationless) structure theorem in [\ref{wigth}].} the suggestive
formula $\left(  \ref{suggestive}\right)  $ suffices. Let $\left\{  \left|
f_{i}^{+}\right\rangle \right\}  $ and $\left\{  \left|  f_{j}^{-}%
\right\rangle \right\}  $ be orthonormal bases of the charged subspaces
$\frak{F}^{\left(  +\right)  }$ and $\frak{F}^{\left(  -\right)  }$ of
$\frak{F}$. Since $U_{S}e^{-\beta H}$ maps $\frak{F}^{\left(  \pm\right)  }$
to $\frak{F}^{\left(  \mp\right)  }$,
\begin{align*}
\operatorname*{Tr}\limits_{\frak{F}}U_{S}e^{-\beta H} &  =\sum_{i,j}%
\left\langle f_{i}^{+}\right|  U_{S}e^{-\beta H}\left|  f_{j}^{-}\right\rangle
\left\langle f_{j}^{-}\right|  U_{S}e^{-\beta H}\left|  f_{i}^{+}\right\rangle
=\sum_{i}\left\langle f_{i}^{+}\right|  U_{S^{2}}e^{-2\beta H}\left|
f_{i}^{+}\right\rangle \\
&  =\operatorname*{Tr}\limits_{\frak{F}^{\left(  +\right)  }}U_{S^{2}%
}e^{-2\beta H}=\operatorname*{Tr}\limits_{\frak{F}^{\left(  -\right)  }%
}U_{S^{2}}e^{-2\beta H}\text{.}%
\end{align*}
But $S^{2}$ is unitary, so
\begin{equation}
\operatorname*{Tr}\limits_{\frak{F}^{\left(  +\right)  }}U_{S^{2}}e^{-2\beta
H}=\prod_{k}\frac{1}{1-\rho_{k}e^{-2\beta\omega_{k}}}\text{,}\label{nmr}%
\end{equation}
where $\left\{  \rho_{k},\omega_{k}\right\}  $ are the joint eigenvalues
(counting multiplicity) of $\left(  S^{2},\Omega\right)  $. Since $\left[
S,S^{2}\right]  =0$, the nonreal $\rho_{k}$ come in conjugate pairs, so both
sides of $\left(  \ref{nmr}\right)  $ are nonnegative. Hence
\[
\operatorname*{Tr}\limits_{\frak{F}}U_{S}e^{-\beta H}=\sqrt{\operatorname*{Tr}%
\limits_{\frak{F}^{\left(  +\right)  }}U_{S^{2}}e^{-2\beta H}\times
\operatorname*{Tr}\limits_{\frak{F}^{\left(  -\right)  }}U_{S^{2}}e^{-2\beta
H}}\text{,}%
\]
proving $\left(  \ref{suggestive}\right)  $.\hfill$\blacksquare$

\section{The Twisted Pair Correlation Function\label{pcsect}}

We study the pair-correlation function, defined for unitary $S$ in definition
\ref{defcintro}. The twisted pair correlation is often written in the
suggestive notation
\begin{equation}
C\left(  t,\bar{f};s,g\right)  =\int\limits_{X\times X}C\left(
t,x;s,y\right)  \bar{f}\left(  x\right)  g\left(  y\right)
dx\,dy\text{,\label{standardc}}%
\end{equation}
where
\begin{equation}
C\left(  t,x;s,y\right)  =\frac{1}{Z_{U_{S}}}\operatorname*{Tr}%
\limits_{\frak{F}}\left[  \left(  \varphi\left(  t,x\right)  \bar{\varphi
}\left(  s,y\right)  \right)  _{+}U_{S}e^{-\beta H}\right]
\text{.\label{standardc2}}%
\end{equation}
Here $C\left(  t,x;s,y\right)  $ is not a function, but is only symbolic
expression similar to the expression $\varphi_{RT}\left(
t,x\right)  $ introduced in \S\ref{basismethod}. Note that the trace operation
in $\left(  \ref{standardc2}\right)  $ is always assumed to be interchanged
with the integral in $\left(  \ref{standardc}\right)  $.

\subsection{\smallskip\label{formalsection}The Integral Kernel $\mathbf{C}%
\left(  t,x;s,y\right)  $}

We begin with a suggestive argument that
\begin{equation}
\int_{0}^{\beta}ds\int_{X}dy\;C\left(  t,x;s,y\right)  g\left(  s,y\right)
=\left(  -D^{2}+\Omega_{x}^{2}\right)  ^{-1}g\left(  t,x\right) \label{conjeq}%
\end{equation}
for smooth functions $g\in\mathcal{T}_{\beta}$ satisfying the periodic
boundary conditions $\left(  \ref{funboundry}\right)  $ for $D$. This
calculation is justified in the remainder of \S\ref{pcsect}.

The field $\bar{\varphi}$ satisfies the analog of the imaginary-time
Klein-Gordon equation,
\begin{equation}
\left(  -\partial_{s}^{2}+\bar{\Omega}_{y}^{2}\right)  \bar{\varphi}\left(
s,y\right)  =0\text{.}\label{kgeqnn}%
\end{equation}
Using this we get an equation for the imaginary-time Feynman Green's
function,
\begin{equation}
\left(  -\partial_{s}^{2}+\bar{\Omega}_{y}^{2}\right)  \left(  \varphi\left(
t,x\right)  \bar{\varphi}\left(  s,y\right)  \right)  _{+}=\delta_{t-s}%
\delta_{x,y}\text{,}\label{propeqn}%
\end{equation}
where $\delta_{x,y}$ is the Dirac measure
\[
\int_{X\times X}\bar{f}\left(  x\right)  g\left(  y\right)  \delta
_{x,y}\,dx\,dy=\left\langle f,g\right\rangle _{\mathcal{E}}\text{.}%
\]

Integrate by parts, interchange the trace and $\left(  -\partial_{s}%
^{2}+\Omega_{y}^{2}\right)  $, and an apply$\left(  \ref{propeqn}\right)  $,
to obtain
\begin{align}
&  \int_{0}^{\beta}ds\int_{X}dy\;C\left(  t,x;s,y\right)  \left(
-\partial_{s}^{2}+\Omega_{y}^{2}\right)  g\left(  s,y\right) \nonumber\\
&  =g\left(  t,x\right)  -\left.  \int_{X}dy\;C\left(  t,x;s,y\right)
\partial_{s}g\left(  s,y\right)  \right|  _{s=0}^{\beta}+\left.  \int
_{X}dy\;\left(  \partial_{s}C\left(  t,x;s,y\right)  \right)  g\left(
s,y\right)  \right|  _{s=0}^{\beta}\label{formalboundry}%
\end{align}
Using the definitions of $\varphi$ and $U_{S}$ for $S$ unitary, and by
cyclicity of the trace,
\begin{align*}
\int_{X}dy\;C\left(  t,x;\beta,y\right)  \partial_{s}g\left(  \beta,y\right)
&  =\frac{1}{Z}\operatorname*{Tr}\left[  \;\varphi\left(  t,x\right)
\bar{\varphi}\left(  \beta,\partial_{s}g\left(  \beta,\cdot\right)  \right)
U_{S}e^{-\beta H}\right] \\
&  =\frac{1}{Z}\operatorname*{Tr}\left[  \varphi\left(  t,x\right)
U_{S}e^{-\beta H}\bar{\varphi}\left(  0,\partial_{s}S^{\ast}g\left(
\beta,\cdot\right)  \right)  \right] \\
&  =\frac{1}{Z}\operatorname*{Tr}\left[  \bar{\varphi}\left(  0,\partial
_{s}S^{\ast}g\left(  \beta,\cdot\right)  \right)  \varphi\left(  t,x\right)
U_{S}e^{-\beta H}\right] \\
&  =\int_{X}dy\;C\left(  t,x;0,y\right)  \partial_{s}S_{y}^{\ast}g\left(
\beta,y\right)  \text{.}%
\end{align*}
The second term in $\left(  \ref{formalboundry}\right)  $ vanishes by applying
the boundary condition $\left(  \ref{funboundry}\right)  $ on $g$. Similarly,
the third term also vanishes, and hence
\[
\int_{0}^{\beta}ds\int_{X}dy\;C\left(  t,x;s,y\right)  \left(  -\partial
_{s}^{2}+\Omega_{y}^{2}\right)  g\left(  s,y\right)  =g\left(  t,x\right)
\text{,}%
\]
suggesting that
\[
\int_{0}^{\beta}ds\int_{X}dy\;C\left(  t,x;s,y\right)  g\left(  s,y\right)
=\left(  -\partial_{t}^{2}+\Omega_{x}^{2}\right)  ^{-1}g\left(  t,x\right)
\text{,}%
\]
as desired.

In the rest of this section, we make precise and justify the above manipulations.

\subsection{Preliminary Estimates and Decomposition of $C_{\beta}$}

We need an estimate to show that $C_{\beta}$ is well-defined and bounded:

\begin{lemma}
\label{eascor}Let $\Omega$ be an admissible classical frequency operator, and
let $\beta>0$. Then for any $n\in\mathbb{Z}^{+}$, $t_{1},...,t_{n}\in\left[
0,\beta\right]  $, and $f_{1},...,f_{n}\in\mathcal{E}$ the time-ordered
product
\[
\left(  \varphi^{\natural}\left(  t_{1},f_{1}^{\natural}\right)
...\varphi^{\natural}\left(  t_{n},f_{n}^{\natural}\right)  \right)
_{+}e^{-\beta H}\;\text{,}%
\]
where the $\natural$'s stand for the independent presence or absence of a bar,
extends to a unique trace-class operator. Furthermore, for each such $n$ and
$\beta$ there exists a constant $K_{\beta,n}$ such that
\begin{equation}
\operatorname*{tr}\left|  \left(  \varphi^{\natural}\left(  t_{1}%
,f_{1}^{\natural}\right)  ...\varphi^{\natural}\left(  t_{n},f_{n}^{\natural
}\right)  \right)  _{+}e^{-\beta H}\right|  <K_{\beta,n}\prod_{i=1}%
^{n}\left\|  \Omega^{-1/2}f_{i}\right\|  _{\mathcal{E}}\label{uniftrn}%
\end{equation}
for all $t_{1},...,t_{n}\in\left[  0,\beta\right]  $.
\end{lemma}%

%TCIMACRO{\TeXButton{Proof}{\proof}}%
%BeginExpansion
\proof
%EndExpansion
We note that $e^{-\alpha H}$ for $\alpha>0$ maps $\frak{F}$ into the domain of
$\sqrt{N}$, which is contained in the any time-ordered product of
imaginary-time fields. Hence expression $\left(  \ref{uniftrn}\right)  $ is
certainly well-defined if all the $t_{k}$ are less than $\beta$.

By equations $\left(  \ref{gdinam1}-\ref{gdinam2},\ref{pheima}-\ref{pheimb}%
\right)  $ and the trace-norm Minkowski inequality, we need only consider
terms of the form
\begin{equation}
f\left(  a_{1},...,a_{n+1}\right)  =e^{-a_{1}H}A_{\pm}^{\#}\left(
g_{1}^{\natural}\right)  e^{-a_{2}H}A_{\pm}^{\#}\left(  g_{2}^{\natural
}\right)  ...A_{\pm}^{\#}\left(  g_{n}^{\natural}\right)  e^{-a_{n+1}%
H}\text{,}\label{nofields}%
\end{equation}
where $g_{i}=\Omega^{-1/2}f_{i}$, $\sum_{i=1}^{n+1}a_{i}=\beta>0$, each
$a_{i}\geq0$, and where $\#$ indicates the presence or absence of a $\ast$.
For simplicity, we bound $\left(  \ref{nofields}\right)  $ in the case that
all the $A_{\pm}^{\#}$ are $A_{+}^{\#}$.

Define the linear functionals $B^{\ast}:$ $\mathcal{E}^{\ast}\rightarrow
B\left(  \frak{F}\right)  $ and $B:\mathcal{E}\rightarrow B\left(
\frak{F}\right)  $ by
\begin{align*}
B^{\ast}\left(  \bar{g}\right)   &  =A_{+}^{\ast}\left(  \bar{g}\right)
\left(  N_{+}+1\right)  ^{-1/2}\\
B\left(  g\right)   &  =\left(  B^{\ast}\left(  \bar{g}\right)  \right)
^{\ast}\text{,}%
\end{align*}
where $N_{+}=\sum_{k}a_{+}^{\ast}\left(  k\right)  a_{+}\left(  k\right)  $.
Then for any $g\in\mathcal{E}$ and any function $h:\mathbb{Z}\rightarrow
\mathbb{C}$%
\begin{align}
\left\|  B^{\#}\left(  g^{\natural}\right)  \right\|  _{\frak{F}} &
\leq\left\|  g\right\|  _{\mathcal{E}}\label{normjoncreate}\\
h\left(  N_{+}\right)  B^{\ast}\left(  \bar{g}\right)   &  =B^{\ast}\left(
\bar{g}\right)  h\left(  N_{+}+1\right) \label{wimpcreate1}\\
h\left(  N_{+}+1\right)  B\left(  g\right)   &  =B\left(  g\right)  h\left(
N_{+}\right)  \text{.}\label{wimpcreate3}%
\end{align}
Temporarily fix the values of the $a_{i}$, and pick $a_{j}\geq\beta/\left(
n+1\right)  $. Consider equation $\left(  \ref{nofields}\right)  $ in terms of
the $B$ and $B^{\ast}$ operators. Using $\left(  \ref{wimpcreate1}\right)
-\left(  \ref{wimpcreate3}\right)  $, to put the factors $\sqrt{N_{+}+s}$ all
next to $\exp\left(  -a_{j}H\right)  $, we have
\begin{equation}
f=e^{-a_{1}H}B^{\#}\left(  g_{1}^{\natural}\right)  e^{-a_{2}H}B^{\#}\left(
g_{2}^{\natural}\right)  ...\left(  \sqrt{P\left(  N_{+}\right)  }%
e^{-a_{j}H/2}\right)  e^{-a_{j}H/2}...B^{\#}\left(  g_{n}^{\natural}\right)
e^{-a_{n+1}H}\label{red}%
\end{equation}
where $P$ is a degree-$n$ polynomial satisfying
\begin{equation}
\left|  P\left(  x\right)  \right|  \leq\left(  x+n+1\right)  ^{n}\text{ for
}x\geq0\label{redpbound}%
\end{equation}
F{}rom the inequality $H\geq\mu N_{+}$, we get
\begin{equation}
\left\|  P\left(  N_{+}\right)  e^{-a_{j}H/2}\right\|  \leq\sup_{x\geq
0}\left(  x+n+1\right)  ^{n}e^{-\mu x}<\infty\text{.}\label{redpap}%
\end{equation}
The existence and uniqueness of a bounded extension in the case $a_{n+1}=0$ is
now clear from $\left(  \ref{red}\right)  $. Then equation $\left(
\ref{red}\right)  $ expresses $f\left(  a_{1},...,a_{n}\right)  $ as a product
of $e^{-a_{j}H/2}$ with many bounded operators. Applying equations $\left(
\ref{normjoncreate}\right)  $, $\left(  \ref{red}\right)  $, and $\left(
\ref{redpap}\right)  $ and the choice of $j$ gives
\[
\operatorname*{tr}\left|  f\left(  a_{1},...,a_{n+1}\right)  \right|
\leq\operatorname*{tr}\left|  e^{-\frac{\beta H}{2n+2}}\right|  \times\left(
\sup_{x\geq0}\left(  x+n+1\right)  ^{n}e^{-\mu x}\right)  \prod_{i=1}%
^{n}\left\|  \Omega^{-1/2}f_{i}\right\|  \text{.}%
\]
But $\exp\left(  -\beta H/\left(  2n+2\right)  \right)  $ is trace-class by
Lemma $\ref{admisbound}$. The $a_{i}$ were arbitrary, so $\left(
\ref{uniftrn}\right)  $ is proved.\hfill$\blacksquare$

We now have

\begin{theorem}
\label{unifcbound}$C_{\beta}:\mathcal{T}_{\beta}\rightarrow\mathcal{T}_{\beta
}$ is well-defined, bounded, and self-adjoint.
\end{theorem}%

%TCIMACRO{\TeXButton{Proof}{\proof}}%
%BeginExpansion
\proof
%EndExpansion
Let $f,g:\left[  0,\beta\right)  \rightarrow\mathcal{E}$ be in $\mathcal{T}%
_{\beta}$. By Lemma \ref{eascor} and Schwarz's inequality for $L^{2}\left(
0,\beta\right)  $,
\[
\left|  \int_{0}^{\beta}\int_{0}^{\beta}\operatorname*{tr}\left(
\varphi\left(  t,\bar{f}\left(  t\right)  \right)  \bar{\varphi}\left(
s,g\left(  s\right)  \right)  \right)  _{+}U_{S}e^{-\beta H}\,dt\,ds\right|
\leq\beta K_{\beta,2}\left\|  \Omega^{-1/2}f\right\|  _{\mathcal{T}_{\beta}%
}\left\|  \Omega^{-1/2}g\right\|  _{\mathcal{T}_{\beta}}\text{.}%
\]
Hence $C_{\beta}$ is well-defined, exists, and is bounded by the Riesz
representation theorem. The self-adjointness of $C_{\beta}$ is an immediate
consequence of $TC$ symmetry (Lemma \ref{ctlemma}).\hfill$\blacksquare$

$C_{\beta}$ behaves nicely under direct sum decompositions:

\begin{lemma}
\label{directsumlemma}Let $\Omega$ be a classical frequency operator of an
admissible Lagrangian with a unitary Lagrangian symmetry $S$. Let the
classical space $\mathcal{E}$ be decomposed into a direct sum of invariant
subspaces of $\Omega$ and $S$,
\[%
\begin{array}
[c]{ccccc}%
\mathcal{E}=\mathcal{E}_{1}\oplus\mathcal{E}_{2}\oplus... &  & \Omega
=\Omega_{1}\oplus\Omega_{2}\oplus... &  & S=S_{1}\oplus S_{2}\oplus...
\end{array}
\]
Let $\mathcal{T}_{\beta,j}=L^{2}\left[  0,\beta\right)  \otimes\mathcal{E}%
_{j}$. Then $C_{\beta}$ also decomposes into a direct sum:
\[
C_{\beta}=C_{\beta,1}\oplus C_{\beta,2}\oplus...
\]
where $C_{\beta,j}:\mathcal{T}_{\beta,j}\rightarrow\mathcal{T}_{\beta,j}$ is
the $S_{j}$-twisted pair correlation operator of the free Bosonic theory with
classical frequency operator $\Omega_{j}$.
\end{lemma}

The proof is straightforward.\pagebreak 

\subsection{\bigskip Rigorous Characterization of $\mathbf{C}_{\beta}$}

We need three more technical lemmas. The first concerns inverses of possibly
unbounded self-adjoint operators:

\begin{lemma}
\label{shortcut}Let $A$ and $B$ be self-adjoint operators on a Hilbert space
$\mathcal{H}$ with $B$ bounded so that
\[
BA=\left.  1\right|  _{\mathcal{\frak{D}}\left(  A\right)  }\text{.}%
\]
Then $B$ maps $\mathcal{H}$ into $\frak{D}\left(  A\right)  $ and
\[
AB=1=\left.  1\right|  _{\mathcal{H}}\text{.}%
\]
\end{lemma}%

%TCIMACRO{\TeXButton{Proof}{\proof}}%
%BeginExpansion
\proof
%EndExpansion
We would like to say $\left(  AB\right)  ^{\ast}=B^{\ast}A^{\ast}=BA=1$, but
since $A$ may be unbounded we must be careful about domains. For $u\in
\frak{D}\left(  A\right)  $ and $x\in\mathcal{H}$,
\[
u\mapsto\left\langle Au,Bx\right\rangle =\left\langle BAu,x\right\rangle
=\left\langle u,x\right\rangle \text{.}%
\]
is a bounded function of $u$. Hence $Bx\in\frak{D}\left(  A^{\ast}\right)
=\frak{D}\left(  A\right)  $ and
\[
\left\langle u,ABx\right\rangle =\left\langle u,x\right\rangle \text{.}%
\]
Since $\frak{\frak{D}}\left(  A\right)  $ is dense,
\[
ABx=x\text{,}%
\]
and so $AB=1$.\hfill{$\blacksquare$}

\begin{definition}
Let $\mathcal{H}$ be a Hilbert space, and let $X$ be a measure space. An
operator-valued function $A:X\rightarrow B\left(  \mathcal{H}\right)  $ is
\textbf{weakly measurable} if the function $\left(  v,f\left(  x\right)
w\right)  $ is a measurable function of $x$ for each $v,w\in\mathcal{H}$. The
integral of such a function is defined by
\[
\left\langle v,\int A\left(  x\right)  w\,dx\right\rangle =\int\left\langle
v,A\left(  x\right)  w\right\rangle \,dx
\]
for all $v,w\in\mathcal{H}$.
\end{definition}

\begin{lemma}
[Semi-noncommutative Fubini Theorem]\label{fubini}Let $X$ be a measure space,
$\mathcal{H}$ be a Hilbert space, and $A:X\rightarrow B\left(  \mathcal{H}%
\right)  $ be a weakly measurable function. If
\[
\int_{X}\operatorname*{Tr}\left|  A\left(  x\right)  \right|  \,dx<\infty
\]
then
\begin{equation}
\int_{X}\operatorname*{Tr}A\left(  x\right)  \,dx=\operatorname*{Tr}\int
_{X}A\left(  x\right)  \,dx\text{.}\label{fubeq}%
\end{equation}
\end{lemma}%

%TCIMACRO{\TeXButton{Proof}{\proof}}%
%BeginExpansion
\proof
%EndExpansion
Let $\left\{  e_{k}\right\}  $ be an arbitrary basis of $\mathcal{H}$. Then
the inequality
\[
\int_{X}\sum_{k}\left|  \left\langle e_{k},A\left(  x\right)  e_{k}%
\right\rangle \right|  \,dx\leq4\int_{X}\operatorname*{Tr}\left|  A\left(
x\right)  \right|  \,dx
\]
follows from the decomposition of $A\left(  x\right)  $ as a linear
combination of positive operators, all of which have trace norm
$\operatorname*{Tr}\left|  \cdot\right|  $ less than or equal to
$\operatorname*{Tr}\left|  A\left(  x\right)  \right|  $:
\[
A\left(  x\right)  =\left(  \operatorname{Re}A\left(  x\right)  \right)
_{+}+\left(  \operatorname{Re}\,A\left(  x\right)  \right)  _{-}+i\left(
\operatorname{Im}\,A\left(  x\right)  \right)  _{+}+i\left(  \operatorname{Im}%
\,A\left(  x\right)  \right)  _{-}\text{.}%
\]
Here $\operatorname{Re}\left(  A\right)  =\frac{1}{2}\left(  A+A^{\ast
}\right)  $, $\operatorname{Im}\left(  A\right)  =\frac{1}{2i}\left(
A-A^{\ast}\right)  $, and $B_{\pm}=\frac{1}{2}\left(  B\pm\left|  B\right|
\right)  $. Equation $\left(  \ref{fubeq}\right)  $ follows by Fubini's
theorem, where the summation over $k $ is considered to be an abstract
Lebesgue integral in the counting measure.\hfill{$\blacksquare$}

\begin{lemma}
\label{killboundaryterm}Let $\Omega$ be admissible and let $t\in\left[
0,\beta\right)  $. Then $\varphi\left(  t,\bar{f}\,\right)  \bar{\varphi
}\left(  \beta,g\right)  U_{S}e^{-\beta H}$ has a unique bounded extension,
which is trace-class and satisfies
\[
\operatorname*{tr}\left(  \varphi\left(  t,\bar{f}\,\right)  \bar{\varphi
}\left(  \beta,g\right)  U_{S}e^{-\beta H}\right)  =\operatorname*{tr}\left(
\bar{\varphi}\left(  0,S^{\ast}g\right)  \varphi\left(  t,\bar{f}\,\right)
U_{S}e^{-\beta H}\right)  \text{.}%
\]
\end{lemma}%

%TCIMACRO{\TeXButton{Proof}{\proof}}%
%BeginExpansion
\proof
%EndExpansion
By Lemma \ref{eascor}, $\varphi\left(  t,\bar{f}\,\right)  \bar{\varphi
}\left(  \beta,g\right)  U_{S}e^{-\beta H}$ has a unique bounded extension,
which is trace-class. Writing
\[
\varphi\left(  t,\bar{f}\,\right)  \bar{\varphi}\left(  \beta,g\right)
U_{S}e^{-\beta H}=\varphi\left(  t,\bar{f}\,\right)  U_{S}e^{-\beta H/2}\times
e^{-\beta H/2}\bar{\varphi}\left(  0,S^{\ast}g\right)  \text{,}%
\]
we notice that both factors extend to trace-class operators by Lemma
\ref{eascor}. By a double-application of cyclicity of the trace,
\begin{align}
\operatorname*{tr}\left(  \varphi\left(  t,\bar{f}\,\right)  \bar{\varphi
}\left(  \beta,g\right)  U_{S}e^{-\beta H}\right)   &  =\operatorname*{tr}%
\left(  e^{-\beta H/2}\bar{\varphi}\left(  0,S^{\ast}g\right)  \varphi\left(
t,\bar{f}\,\right)  U_{S}e^{-\beta H/2}\right) \nonumber\\
&  =\operatorname*{tr}\left(  \bar{\varphi}\left(  0,S^{\ast}g\right)
\varphi\left(  t,\bar{f}\,\right)  U_{S}e^{-\beta H}\right)  \text{.}%
\tag*{$\blacksquare$}%
\end{align}

We may now make rigorous the argument of section \ref{formalsection}.

\begin{theorem}
\label{tppctheorem}Let $S$ be a unitary Lagrangian symmetry of an admissible
free Lagrangian $\mathcal{L}$. Then $C_{\beta}=\left(  -D^{2}+\Omega
^{2}\right)  _{\mathcal{T}_{\beta}}^{-1}$, where $\Omega$ is identified with
$1\otimes\Omega:\mathcal{T}_{\beta}\rightarrow\mathcal{T}_{\beta}$.
\end{theorem}%

%TCIMACRO{\TeXButton{Proof}{\proof}}%
%BeginExpansion
\proof
%EndExpansion
We claim that we only need consider the case that $\mathcal{E}=\mathbb{C}$.
Since $\left[  S,\Omega\right]  =0$, we may choose a basis $\left\{
e_{k}\right\}  $ of $\mathcal{E}$ of simultaneous eigenvectors of $S$ and
$\Omega$. Then $\left(  -D^{2}+\Omega^{2}\right)  ^{-1}$ is reduced by the
direct sum
\[
\mathcal{T}_{\beta}=\oplus_{k}L^{2}\left[  0,\beta\right)  \otimes
\operatorname*{Span}\left(  e_{k}\right)  \text{.}%
\]
By Lemma \ref{directsumlemma}, $C_{\beta}$ is also reduced, proving our claim.

By Lemma $\ref{shortcut}$, all we have to show is that
\begin{equation}
\left\langle f,C_{\beta}\left(  -D^{2}+\Omega^{2}\right)  g\right\rangle
_{\mathcal{T}_{\beta}}=\left\langle f,g\right\rangle _{\mathcal{T}_{\beta}%
}\label{lasttodo}%
\end{equation}
for $g$ in the domain of $-D^{2}+\Omega^{2}$. By standard Sobolev space
results (or Lebesgue's density theorem), such $g$ may be represented by a
function which is twice-differentiable almost everywhere and satisfies
\begin{align}
g\left(  \beta\right)   &  =Sg\left(  0\right) \nonumber\\
g^{\prime}\left(  \beta\right)   &  =Sg^{\prime}\left(  0\right) \nonumber\\
g^{\prime}\left(  b\right)  -g^{\prime}\left(  a\right)   &  =\int_{a}%
^{b}g^{\prime\prime}\left(  x\right)  \,dx\text{, \ \ }0\leq a\leq b\leq
\beta\text{,}\label{h11}%
\end{align}
where $S$ is now just a complex number and $Dg=g^{\prime}$. For $E,F\in
\frak{F}_{0}$, we have the identity
\[
\left\langle E\right|  \,\bar{\varphi}\left(  s,\left(  -\frac{d^{2}}{ds^{2}%
}+\Omega^{2}\right)  g\left(  s\right)  \right)  \,\left|  F\right\rangle
=\frac{d}{ds}\left\langle E\right|  \left(  -\bar{\varphi}\left(  s,g^{\prime
}\left(  s\right)  \right)  +\frac{\partial\bar{\varphi}}{\partial s}\left(
s,g\left(  s\right)  \right)  \right)  \,\left|  F\right\rangle \text{.}%
\]
Let $\left\{  E_{n}\right\}  \subseteq\frak{F}$ be a basis of eigenfunctions
of $N$. We compute%

\begin{align}
&  \left(  f,C_{\beta}\,\left(  -D^{2}+\Omega^{2}\right)  g\right) \nonumber\\
&  =\frac{1}{Z}\sum_{n}\int_{0}^{\beta}\int_{0}^{t}dt\,ds\;\frac{d}%
{ds}\left\langle E_{n}\right|  \left(  -\bar{\varphi}\left(  s,\frac{dg}%
{ds}\right)  +\frac{\partial\bar{\varphi}}{\partial s}\left(  s,g\right)
\right)  \varphi\left(  t,\bar{f}\,\left(  t\right)  \right)  U_{S}e^{-\beta
H}\left|  E_{n}\right\rangle \nonumber\\
&  \;\;\;\;+\frac{1}{Z}\sum_{n}\int_{0}^{\beta}\int_{t}^{\beta}dt\,ds\;\frac
{d}{ds}\left\langle E_{n}\right|  \varphi\left(  t,\bar{f}\left(  t\right)
\right)  \left(  -\bar{\varphi}\left(  s,\frac{dg}{ds}\right)  +\frac
{\partial\bar{\varphi}}{\partial s}\left(  s,g\right)  \right)  U_{S}e^{-\beta
H}\left|  E_{n}\right\rangle \nonumber\\
&  =\frac{1}{Z}\sum_{n}\int_{0}^{\beta}dt\,\;\left\langle E_{n}\right|
\left[  -\bar{\varphi}\left(  t,\frac{dg}{dt}\right)  +\frac{\partial
\bar{\varphi}}{\partial t}\left(  t,g\right)  ,\varphi\left(  t,\bar{f}\left(
t\right)  \right)  \right]  U_{S}e^{-\beta H}\left|  E_{n}\right\rangle
+BT\label{useh11}\\
&  =\frac{1}{Z}\sum_{n}\int_{0}^{\beta}dt\,\;\left(  f\left(  t\right)
,g\left(  t\right)  \right)  _{\mathcal{E}}\,\left\langle E_{n}\right|
U_{S}e^{-\beta H}\left|  E_{n}\right\rangle +BT\nonumber\\
&  =\left(  f,g\right)  _{\mathcal{T}_{\beta}}+BT\text{,}\nonumber
\end{align}
\newline where $BT$ stands for the boundary terms. We were able to move the
integrations inside of the trace using Lemma $\ref{fubini}$ and the estimate
of Lemma \ref{eascor}. Equation $\left(  \ref{useh11}\right)  $ used $\left(
\ref{h11}\right)  $.

We consider the boundary terms:
\begin{align*}
BT &  =-\frac{1}{Z}\sum_{n}\int_{0}^{\beta}dt\;\left\langle E_{n}\right|
\left(  -\bar{\varphi}\left(  0,D_{s}g\right)  +\frac{\partial\bar{\varphi}%
}{\partial s}\left(  0,g\right)  \right)  \varphi\left(  t,\bar{f}\left(
t\right)  \right)  U_{S}e^{-\beta H}\left|  E_{n}\right\rangle \\
&  \;\;\;\;\,+\frac{1}{Z}\sum_{n}\int_{0}^{\beta}dt\,\;\left\langle
E_{n}\right|  \varphi\left(  t,\bar{f}\left(  t\right)  \right)  \left(
-\bar{\varphi}\left(  \beta,D_{s}g\right)  +\frac{\partial\bar{\varphi}%
}{\partial s}\left(  \beta,g\right)  \right)  U_{S}e^{-\beta H}\left|
E_{n}\right\rangle \\
&  =\frac{1}{Z}\int_{0}^{\beta}dt\;\operatorname*{Tr}\bar{\varphi}\left(
0,D_{s}g\left(  0\right)  \right)  \varphi\left(  t,\bar{f}\left(  t\right)
\right)  U_{S}e^{-\beta H}\\
&  \;\;\;\;\,-\frac{1}{Z}\int_{0}^{\beta}dt\;\operatorname*{Tr}\varphi\left(
t,\bar{f}\left(  t\right)  \right)  \bar{\varphi}\left(  \beta,D_{s}g\left(
\beta\right)  \right)  U_{S}e^{-\beta H}\\
&  \;\;\;\;\,+\frac{1}{Z}\int_{0}^{\beta}dt\,\;\operatorname*{Tr}%
\varphi\left(  t,\bar{f}\left(  t\right)  \right)  \frac{\partial\bar{\varphi
}}{\partial s}\left(  \beta,g\left(  \beta\right)  \right)  U_{S}e^{-\beta
H}\\
&  \;\;\;\;\,-\frac{1}{Z}\int_{0}^{\beta}dt\;\operatorname*{Tr}\frac
{\partial\bar{\varphi}}{\partial s}\left(  0,g\left(  0\right)  \right)
\varphi\left(  t,\bar{f}\left(  t\right)  \right)  U_{S}e^{-\beta H}%
\end{align*}
We were able to interchange integration and the trace for the same reasons as
above. The first two terms cancel by Lemma \ref{killboundaryterm}. The last
two cancel similarly.\hfill{$\blacksquare$}\pagebreak 

\section{The Antiunitary Case, Real Fields}

We would like to prove an analog of Theorem $\ref{tppctheorem}$ for
antiunitary classical symmetries, as well as a theorem for symmetries of real
scalar fields. Given that we have not required the choice of an arbitrary
conjugation on our classical space $\mathcal{E}$,\footnote{All use of the
arbitary representation $\mathcal{E}=L^{2}\left(  X\right)  $ was for
notational purposes only.} it is surprising that unification of the unitary
and antiunitary cases results from consideration of the real scalar field.

\subsection{The Extended Pair Correlation Operator}

We note that if $V:\mathcal{E}\rightarrow\mathcal{E}$ is antiunitary then in
general
\[
\operatorname*{tr}\left[  \left(  \varphi\left(  t,x\right)  \varphi\left(
s,y\right)  \right)  _{+}U_{V}e^{-\beta H}\right]  \neq0\text{.}%
\]
Hence the important operator for Wick's theorem is no longer the pair
correlation operator $C_{\beta}$. We define

\begin{definition}
Let $\Omega$ be a classical frequency operator on $\mathcal{E}$ with
antiunitary classical symmetry $V$. We define the \textbf{extended space of
classical fields} $\mathbb{E}=\mathcal{E}^{\ast}\oplus\mathcal{E}$. The
\textbf{extended path space }is
\[
\mathbb{T}_{\beta}=L^{2}\left(  0,\beta\right)  \otimes\mathbb{E}\text{,}%
\]
and the \textbf{extended twisted pair correlation operator} $\hat{C}_{\beta
}:\mathbb{T}_{\beta}\rightarrow\mathbb{T}_{\beta}$ is the operator which
satisfies
\begin{align*}
\left(  \bar{f}\left(  t\right)  \oplus g\left(  t\right)  ,\hat{C}\;\bar
{h}\left(  t\right)  \oplus k\left(  t\right)  \right)  _{\mathbb{T}_{\beta}}
&  =\frac{1}{Z_{\beta,U_{V}}}\int\int\operatorname*{tr}\left(  \left(
\bar{\varphi}\left(  t,f\left(  t\right)  \right)  \varphi\left(  s,\bar
{h}\left(  s\right)  \right)  \right)  _{+}U_{V}e^{-\beta H}\right)
\;dt\,ds\\
&  \;\;\;\;+\frac{1}{Z_{\beta,U_{V}}}\int\int\operatorname*{tr}\left(  \left(
\bar{\varphi}\left(  t,f\left(  t\right)  \right)  \bar{\varphi}\left(
s,k\left(  s\right)  \right)  \right)  _{+}U_{V}e^{-\beta H}\right)
\;dt\,ds\\
&  \;\;\;\;+\frac{1}{Z_{\beta,U_{V}}}\int\int\operatorname*{tr}\left(  \left(
\varphi\left(  t,\bar{g}\left(  t\right)  \right)  \varphi\left(  s,\bar
{h}\left(  s\right)  \right)  \right)  _{+}U_{V}e^{-\beta H}\right)
\;dt\,ds\\
&  \;\;\;\;+\frac{1}{Z_{\beta,U_{V}}}\int\int\operatorname*{tr}\left(  \left(
\varphi\left(  t,\bar{g}\left(  t\right)  \right)  \bar{\varphi}\left(
s,k\left(  s\right)  \right)  \right)  _{+}U_{V}e^{-\beta H}\right)
\;dt\,ds\text{.}%
\end{align*}
\end{definition}

We note that if the symmetry $V$ were unitary, then the middle two terms would
vanish, reducing consideration to the pair correlation operators associated to
$\left(  \Omega,V\right)  $ and $\left(  \bar{\Omega},\bar{V}\right)  $.

\subsection{\label{realfieldsect}The relationship between real and complex
scalar fields}

We reduce consideration of antiunitary symmetries of a complex scalar field to
consideration of (classically unitary) symmetries of a real scalar field. Had
we required that our space of classical fields $\mathcal{E}$ come equipped
with a conjugation which commuted with $\Omega$\footnote{so that the
Klein-Gordon equation has real solutions} then our complex field theory would
be a direct sum of two real fields.\footnote{We would likewise need to
restrict consideration Lagrangian symmetries which commute with conjugation on
$\mathcal{E}$.} Although we impose no reality condition on $\Omega$ nor
conjugation on $\mathcal{E}$, we will see that in a certain sense our complex
field \textit{is}\textbf{\ }a real field.

\begin{definition}
The \textbf{natural conjugation }$\bar{\cdot}:\mathbb{E}\rightarrow\mathbb{E}$
is given by
\[
\overline{\bar{f}\oplus g}=\bar{g}\oplus f\text{.}%
\]
A \textbf{real operator} $R:\mathbb{E}\rightarrow\mathbb{E}$ is one that
commutes with conjugation. Let $\Omega:\mathcal{E}\rightarrow\mathcal{E}$ be a
classical frequency operator of a complex field $\varphi_{RT}$, as above. Let
$S$ and $V$ be unitary and antiunitary Lagrangian symmetries of $\Omega$,
respectively. Define the \textbf{associated real field }$\psi_{RT}%
:\mathbb{R}\times\mathbb{E}\rightarrow Op\left(  \frak{F}\right)  $ by
\[
\psi_{RT}\left(  t,\bar{f}\oplus g\right)  =\varphi_{RT}\left(  t,\bar
{f}\,\right)  +\varphi_{RT}^{\ast}\left(  t,g\right)  \text{,}%
\]
and the \textbf{associated imaginary-time real field }$\psi:\mathbb{R}%
\times\mathbb{E}\rightarrow Op\left(  \frak{F}\right)  $ by
\[
\psi\left(  t,\bar{f}\oplus g\right)  =e^{-tH}\psi_{RT}\left(  0,\bar{f}\oplus
g\right)  e^{tH}\text{.}%
\]
Furthermore, define
\begin{align*}
\mathbb{\Omega}_{\mathbb{R}} &  =\bar{\Omega}\oplus\Omega\\
\mathbb{S} &  =\bar{S}\oplus S\\
\mathbb{V}\left(  \bar{f}\oplus g\right)   &  =\overline{Vg}\oplus Vf\\
\mathbb{A}^{\ast}\left(  \bar{f}\oplus g\right)   &  =A_{+}^{\ast}\left(
\bar{f}\,\right)  +A_{-}^{\ast}\left(  g\right) \\
\mathbb{A}\left(  \bar{f}\oplus g\right)   &  =A_{-}\left(  \bar{f}\,\right)
+A_{+}\left(  g\right)  =\left(  \mathbb{A}^{\ast}\left(  \overline{\bar
{f}\oplus g}\right)  \right)  ^{\ast}\\
\mathbb{T}_{\beta} &  =L^{2}\left[  0,\beta\right)  \otimes\mathbb{E}%
\end{align*}
and define $\mathbb{D}_{\mathbb{S}}$ and $\mathbb{D}_{\mathbb{V}}%
:\mathbb{T}_{\beta}\rightarrow\mathbb{T}_{\beta}$ analogously to $D$.
\end{definition}

We have the following

\begin{theorem}
$\psi_{RT}$ is a free real scalar field with classical frequency operator
$\mathbb{\Omega}_{\mathbb{R}}$, i.e.

\begin{enumerate}
\item $\mathbb{\Omega}_{\mathbb{R}}$ is real.

\item $\psi_{RT}$ is self-adjoint, i.e. $\left(  \psi_{RT}\left(  t,q\right)
\right)  ^{\ast}=\psi_{RT}\left(  t,\bar{q}\right)  $.

\item $\partial_{t}^{2}\psi_{RT}\left(  t,q\right)  =-\psi_{RT}\left(
t,\mathbb{\Omega}_{\mathbb{R}}^{2}q\right)  $ strongly on $\frak{F}_{0}$ for
$q\in\mathcal{D}\left(  \Omega_{\mathbb{R}}^{2}\right)  $.

\item $\left[  \psi_{RT}\left(  t,q\right)  ,\frac{\partial\psi_{RT}}{\partial
t}\left(  t,r\right)  \right]  =i\left(  \bar{q},r\right)  _{\mathbb{E}}$.

\item $\left[  \psi_{RT}\left(  t,q\right)  ,\psi_{RT}\left(  t,r\right)
\right]  =0$.

\item  Successively applying $\psi_{RT}\left(  t,\cdot\right)  $ and
$\partial_{t}\psi_{RT}\left(  t,\cdot\right)  $, and to $\left|
0\right\rangle $ gives a dense subset of $\frak{F}$.

\noindent Furthermore,

\item $\mathbb{A}^{\ast}$ is the creation functional of $\psi_{RT}$, i.e.
\[
\psi_{RT}\left(  t,q\right)  =\frac{1}{\sqrt{2}}\left(  \mathbb{A}^{\ast
}\left(  \mathbb{\Omega}_{\mathbb{R}}^{-1/2}e^{it\mathbb{\Omega}_{\mathbb{R}}%
}q\right)  +\mathbb{A}\left(  \mathbb{\Omega}_{\mathbb{R}}^{-1/2}%
e^{-it\mathbb{\Omega}_{\mathbb{R}}}q\right)  \right)
\]
and
\[
\left[  \mathbb{A}\left(  q\right)  ,\mathbb{A}^{\ast}\left(  r\right)
\right]  =\left\langle \bar{q},r\right\rangle _{\mathbb{E}}\text{.}%
\]

\item $\mathbb{S}$ and $\mathbb{V}$ are real and unitary, and the real-field
Fock space implementations of $\mathbb{S}$ and $\mathbb{V}$, which satisfy
\begin{align*}
\mathbb{U}_{\mathbb{S}}\left|  0\right\rangle  &  =\mathbb{U}_{\mathbb{V}%
}\left|  0\right\rangle =\left|  0\right\rangle \\
\mathbb{U}_{\mathbb{S}}\psi_{RT}\left(  t,q\right)  \mathbb{U}_{\mathbb{S}%
}^{\ast} &  =\psi_{RT}\left(  t,\mathbb{S}^{\ast}q\right) \\
\mathbb{U}_{\mathbb{V}}\psi_{RT}\left(  t,q\right)  \mathbb{U}_{\mathbb{V}%
}^{\ast} &  =\psi_{RT}\left(  t,\mathbb{V}^{\ast}q\right)  \text{,}%
\end{align*}
are simply given by
\[%
\begin{array}
[c]{ccc}%
\mathbb{U}_{\mathbb{S}}=U_{S} & \text{and} & \mathbb{U}_{\mathbb{V}}%
=U_{V}\text{.}%
\end{array}
\]

\item $\hat{C}$ is the twisted pair correlation operator of $\psi_{RT}$ with
the symmetry $\mathbb{V}$, i.e.
\[
\left(  \bar{f}\oplus g,\hat{C}\,\bar{h}\oplus k\right)  =\int_{0}^{\beta}%
\int_{0}^{\beta}\operatorname*{tr}\left[  \left(  \psi\left(  t,\overline
{\bar{f}\oplus g}\right)  \psi\left(  s,\bar{h}\oplus k\right)  \right)
_{+}U_{V}e^{-\beta H}\right]  \;dt\,ds\text{.}%
\]
\end{enumerate}
\end{theorem}

\noindent We have the following theorem concerning real scalar fields:

\begin{theorem}
\label{toobig}Let$\;\mathcal{\tilde{E}}$ be a Hilbert space with conjugation,
and let $\varphi_{RT}^{\mathbb{R}}:\mathbb{R\times}\mathcal{\tilde{E}%
}\rightarrow\mathbb{C}$ be a free real scalar field with admissible real
classical frequency operator $\mathbb{\tilde{\Omega}}^{\;}:\mathcal{\tilde{E}%
}\rightarrow\mathcal{\tilde{E}}$. Let $\tilde{S}$ be a real unitary Lagrangian
symmetry of $\mathbb{\tilde{\Omega}}$. Then $U_{\tilde{S}}$ is twist-positive
and the corresponding pair-correlation operator $\tilde{C}_{\beta}$ satisfies
\begin{equation}
\tilde{C}_{\beta}=\left(  -\tilde{D}^{2}+\mathbb{\tilde{\Omega}}^{2}\right)
_{\tilde{T}_{\beta}}^{-1}\text{,}\label{finally}%
\end{equation}
where $\tilde{D}$ is defined analogously to $D$, and where $\tilde{\Omega}$ is
identified with $I\otimes\tilde{\Omega}:\tilde{T}_{\beta}\rightarrow\tilde
{T}_{\beta}$.
\end{theorem}%

%TCIMACRO{\TeXButton{Proof}{\proof}}%
%BeginExpansion
\proof
%EndExpansion
Twist positivity is proved by replacing the use of $TC$ symmetry in the proof
of Theorem $\ref{tptheorem}$ with the observation that the nonreal eigenvalues
of the real operator $\tilde{S}$ come in complex conjugate pairs. Equation
$\left(  \ref{finally}\right)  $ may be proved by slight notational changes in
the proof of Theorem \ref{tppctheorem}. \hfill{$\blacksquare$}

The previous two theorems reduce the antiunitary case to a triviality:

\begin{corollary}
\label{antiunitcor}Let $\Omega$ be an admissible classical frequency operator
of a free complex scalar field. Let $V$ be an antiunitary Lagrangian symmetry.
Then the extended pair correlation operator $\hat{C}_{\beta}$ is positive
definite. In particular,
\[
\hat{C}_{\beta}=\left(  -\mathbb{D}_{\mathbb{V}}^{2}+\Omega_{\mathbb{R}}%
^{2}\right)  _{\mathbb{T}_{\beta}}^{-1}\text{,}%
\]
where $\Omega_{\mathbb{R}}$ is identified with $I\otimes\Omega_{\mathbb{R}}$.
\end{corollary}

Furthermore, we note that Theorem \ref{toobig} applies not only applies to
Lagrangian symmetries of complex fields, but in general to symmetries which
mix the subspaces $\mathcal{E}$ and $\mathcal{E}^{\ast}$ of $\mathbb{E}%
$.\footnote{The simplest example is given by $\mathcal{E}=\mathbb{C}^{2}$,
$\tilde{S}\left(  \bar{f}\oplus g\right)  =\left[  \overline{\left(
\sigma_{x}f+s_{z}g\right)  }\oplus\left(  s_{z}f+\sigma_{z}g\right)  \right]
/\sqrt{2}$, where $\sigma_{x}\left(  z_{1},z_{2}\right)  =\left(  z_{2}%
,z_{1}\right)  $, $s_{z}\left(  z_{1},z_{2}\right)  =\left(  \bar{z}_{1}%
,-\bar{z}_{2}\right)  $, $f\mapsto\bar{f}$ is given by definition
\ref{defadmis2}, and $z\mapsto\bar{z}$ is just complex conjugation.} Since the
Fock-space implementations of these additional symmetries will mix particle
and antiparticle states, they are somewhat less natural.

\end{document}